\documentstyle[prb,aps,psfig,multicol,eqsecnum]{revtex}

\newcommand{\crossprod}{\times}
\newcommand{\be}{\begin{equation}}
\newcommand{\ee}{\end{equation}}
\newcommand{\barr}{\begin{eqnarray}}
\newcommand{\earr}{\end{eqnarray}}
\newcommand{\breakeq}{\nonumber \\ &&}
\newcommand{\unit}[1]{\,\mathrm{#1}}

\newcommand{\rvec}{\mbox{${\bf r}$}}

\newcommand{\rpvec}{\mbox{${\bf r}'$}}

\newcommand{\Bvec}{\mbox{${\bf B}$}}
\newcommand{\Evec}{\mbox{${\bf E}$}}
\newcommand{\Lvec}{\mbox{${\bf L}$}}
\newcommand{\Xvec}{\mbox{${\bf X}$}}

\newcommand{\nhat}{\mbox{$\hat{n}'$}}

\begin{document} 
\draft
\title{Light scattering from disordered overlayers of metallic nanoparticles} 
\author{Peter Johansson}
\address{Division of Solid State Theory,  Department of Physics,
   University of Lund, S\"olvegatan 14 A, S-223\,62 Lund, Sweden}
\address{and Department of Natural Sciences,  University of \"Orebro,  
          S-701\,82 \"Orebro, Sweden}

\date{\today}
\maketitle
\begin{abstract}
We develop a theory for light scattering from a disordered layer
of metal nanoparticles resting on a sample. 
Averaging over different disorder realizations is done by a 
coherent potential approximation.  The calculational scheme takes
into account effects of retardation, multipole excitations, and 
interactions with the sample. 
We apply the theory  to a system similar to the one studied experimentally
by Stuart and Hall [Phys.\ Rev.\ Lett.\ {\bf 80}, 5663 (1998)] who used a 
layered Si/SiO$_2$/Si sample.
The calculated results agree rather well with the experimental ones.  In
particular we find conspicuous maxima in the scattering intensity at long
wavelengths (much longer than those corresponding to plasmon resonances in
the particles). We show that these maxima have their origin in interference
phenomena in the layered sample.
\end{abstract}



\begin{multicols}{2}

\section{Introduction}
\label{SecIntro}

During the last several years there has been an increased interest in
near-field optics. This development has been driven in part by the 
development of the near-field scanning optical microscope, 
but also by a growing awareness of the importance of near-field effects 
and the influence they have on the radiated, far fields.
Examples of probes that investigate or make use of the
coupling between near-field and far-field effects include
surface-enhanced Raman scattering (SERS),\cite{SERS1}
light emission from scanning tunneling microscopes,\cite{RB}
transmission through, and reflection from, narrow
gratings,\cite{GratingsE,GratingsT}
and scattering from particle arrays.\cite{SH,Lamp,Lind} 
Applications range from single-molecule 
spectroscopy using SERS to creating microcavity light sources\cite{Cavities}
and using the plasmon resonances  of metallic nanoparticles for 
the transmission of ``optical'' signals at a sub-wavelength length 
scale.\cite{Brongersma}

The optical properties of small metallic particles
and collections thereof have been studied
intensively both experimentally 
\cite{Yam,HH,Leit,SH,Lamp,Lind}
and theoretically.
\cite{Oht,Lamb,PerLi,Wok,Bobb,Fuchs,Stef,Abajo}
Many of these studies have focused on measuring and calculating 
properties such as average transmission and absorption in a colloidal
solutions or surface films, and to determine the role of inter-particle
interactions on these properties.
At present there is also a lot of interest in studying the interaction
between surface plasmons and particles resting on the surface in question
since the presence of the particles can induce surface-plasmon 
band gaps.\cite{Kit,Boz}

In this paper we will focus on, and develop a theory for diffuse
light scattering from disordered 
overlayers of nanoparticles resting on a sample. 
The motivation for this work was initially provided by an experiment in
which Stuart and Hall (SH)\cite{SH} measured the diffuse light
scattering from random  silver nanoparticle arrays fabricated on top of
dielectric as well as semiconducting and metallic samples.
The silver particles had a 
a size of some 100 nm and the intensity of scattered light as a function
of photon energy showed rather intriguing features. 
The most striking result was that the scattering intensity
had pronounced maxima at very long wavelengths ($\lambda > 1000$ nm)
when the sample consisted of Si and SiO$_2$ in a multilayer structure.
Stuart and Hall\cite{SH} attributed this to an enhancement of the 
dipole-dipole interaction between the different particles mediated by
waveguide modes that are supported by the layered sample.
In this scenario, incident light is initially scattered into a waveguide
mode in the sample by the nanoparticle array. The electromagnetic field 
associated with a waveguide mode is in general evanescent outside the
sample. Therefore light scattered into such a mode is ``trapped'' by the sample
and can travel a relatively long distance before it is damped due to losses in
the sample or because it is rescattered into free space by the
nanoparticles. Thus, it is clear that the presence of a sample can lead to
an increase in the particle-particle interactions.
Nevertheless, as we will see below, the results of this calculation do not
support the scenario of Stuart and Hall.

A rigorous theory that can be used to calculate the scattering intensity
from the system in question must include a number of ingredients: 
particle-particle
interactions through near-field and far-field interactions as well as
interactions mediated by the sample have to be taken into account.
In addition, the nanoparticle layer is disordered and the calculation must
deal with the averaging over different disorder realizations.
Here we do this by means of 
a lattice-gas model\cite{PerLi} in which the sites of a regular
lattice are either occupied by spheres
of a certain size and composition or else empty; 
the averaging over different realizations 
of the disorder is then done by a coherent potential approximation 
(CPA).\cite{Rick,PerLi}
In previous work, Persson and Liebsch (PL)\cite{PerLi} dealt with a similar
model within the CPA. Their treatment accounted for non-retarded
dipole-dipole interactions between the particles.
Later Stefanou and Modinos (SM)\cite{Stef} studied the same model. Their
treatment of disorder effects used the simpler average T-matrix approximation
(ATA), however, they did include retardation
effects as well as sample-mediated interactions in their theory.
Both PL and SM concentrated on calculating the average transmission
through, and reflection from arrays of relatively small particles 
($R\alt 10 \unit{nm}$). 
Meier, Wokaun, and Liao\cite{Wok} have also studied
dipole-dipole interactions in a self-consistent way 
in a disordered array of particles by convoluting the response of a
single-particle with a function describing the distribution of particles in
the array. 

Our theory goes beyond earlier treatments\cite{PerLi,Wok,Stef} in that 
it simultaneously includes effects of retardation, multipolar excitations,
interactions with the sample, and disorder. 
Moreover, since we wish to calculate the intensity of diffusely scattered
light, we do not only calculate disorder-averaged fields and particle
polarizabilities (which one obtains
from the CPA), but also disorder-averaged intensities. 
We calculate these intensities within a conserving approximation, thus
using an averaging procedure that includes exactly the same kind of
scattering processes as the CPA calculation of the average fields.
In the language of many-body theory,\cite{Mahan} evaluating the average
polarizabilities is quite analogous to a single-particle self energy
calculation, 
whereas the calculation of the diffuse scattering intensities requires that
a vertex function (involving two-particle correlations) is solved for.

The calculated results show good agreement with the experimental ones.  
In particular, as in the experiment,\cite{SH} 
a series of resonance peaks emerge in the spectrum of diffusely 
scattered light at relatively small photon energies. 
A detailed study of the behind-lying mechanisms shows that these resonances
occur whenever the field that drives the plasma oscillations in the silver
particles reaches a maximum.  This driving field is the sum of a
contribution from the incident wave and a contribution from the wave
reflected from the sample.  The driving field displays a number of
oscillations due to interference between waves reflected from the different
interfaces in the multilayered sample. 
Thus, the interpretation we arrive at is simpler than the one proposed by
SH.\cite{SH} It is, however, consistent with conclusions  drawn from earlier
experimental results by Leitner {\it et al.}\cite{Leit} who studied light
scattering from a silver island film resting on a layered sample and
observed characteristic changes in the scattering spectrum as the sample
geometry was varied.

The rest of the paper is organized in the following way.  In Sec.\
\ref{SecTheory}
the basic theory is outlined, while Sec.\ \ref{SecDisorder} describes the
theory involved in the disorder-averaging.  
In Sec.\ \ref{SecRes} the numerical results are 
presented and discussed, and Sec.\ \ref{SecSummary} gives a brief summary
of the paper.
Three appendices contain information on more technical aspects
of the calculations.

\section{Basic theory}
\label{SecTheory}

The system  we consider consists of an array of spheres placed on 
or above a semi-infinite sample that may be layered.  
A schematic illustration of the system is shown in Fig.\ \ref{Fig1}.
The optical properties of the sample and particle materials are taken into
account through local dielectric functions tabulated by Palik.\cite{Palik}
Since we will deal with relatively large particles non-local corrections to
the dielectric properties should be relatively small.

To begin with we assume that the array of spheres is perfectly 
ordered, but later we will relax this assumption replacing the array
with a lattice gas where a particular
site is occupied by a sphere with probability $p$
and unoccupied with probability $1-p$.

\subsection{Kirchoff integrals}

The  basic task before us 
is to calculate the electric and magnetic fields everywhere in space
given an incident wave  with electric field ${\bf E}^{\rm ext}$,
and magnetic field ${\bf B}^{\rm ext}$.
We will use the vector equivalents of the Kirchoff integral\cite{Jack} 
to do this.
These formulae read (a time-dependence $e^{-i\omega t}$ is implicitly
assumed everywhere) 
\barr
 \Evec(\rvec)=&&
 \Evec^{\rm ext}(\rvec) +
  \int dS'
  \left[
         ikc(\nhat\crossprod\Bvec(\rpvec))G(\rvec,\rpvec)
\right.
\breakeq 
\left.
         + (\nhat\crossprod\Evec(\rpvec))\crossprod\nabla'G
         + (\nhat\cdot\Evec(\rpvec))\nabla'G
                                       \right]
\label{KirchE}
\earr
and 
\barr
 \Bvec(\rvec)=&&
 \Bvec^{\rm ext}(\rvec) +
  \int dS'
  \left[
        - \frac{ik}{c} (\nhat\crossprod\Evec(\rpvec))G(\rvec,\rpvec)
\right.
\breakeq 
\left.
         + (\nhat\crossprod\Bvec(\rpvec))\crossprod\nabla'G
         + (\nhat\cdot\Bvec(\rpvec))\nabla'G
                                       \right].
\label{KirchB}
\earr
Here $G$ denotes the Green's function of the scalar Helmholtz equation in free
space, thus $G$  can be written 
\be
 G (\rvec, \rpvec)= \frac {e^{ik|\rvec-\rpvec|}} { 4\pi |\rvec-\rpvec| }
\ee
and solves
\be
[\nabla^2 + k^2] G(\rvec,\rpvec) = - \delta^{(3)}(\rvec-\rpvec),
\ee
where $k=\omega/c$.
The integrals appearing in Eqs.\ (\ref{KirchE}) and (\ref{KirchB}) 
run over all
surfaces that enclose the nanoparticles as well as the sample, and the 
$\Evec$ and $\Bvec$ fields appearing inside the integrals are the
exact fields at these surfaces. Thus, Eqs.\ (\ref{KirchE}) and 
(\ref{KirchB})
are coupled integral equations from which the electromagnetic field
always, at least in principle, can be calculated. 

The Kirchoff integrals become very useful also for detailed calculations 
once we have analytic 
expressions for the fields inside the spheres in the array and in the sample.
In the rest of this section, however, they  will 
mainly be used as a tool to aid our thinking.

\subsection{Multipole fields and sphere response}

The electromagnetic field inside and around 
each sphere  can be expressed 
in terms of electric (E) and magnetic (M) multipole fields. 
An {\em electric} multipole field is defined by (we use Jackson's
definitions\cite{Jack})
\be 
 \Bvec^{(E)}=\frac{k}{c}z_l(k_rr)\Xvec_{lm}; \ \ \  \ 
 \Evec^{(E)}=\frac{i}{\epsilon_r}\nabla\crossprod
 \left[z_l(k_rr)\Xvec_{lm}\right],
\label{Emultipole}
\ee
where
$k_r$ is the magnitude of the wave vector $k_r = \sqrt{\epsilon_r} \omega/c$
in a material with relative dielectric function $\epsilon_r$.\cite{Palik}
A {\em magnetic} multipole is associated with the fields
\be 
 \Evec^{(M)}=kz_l(k_rr)\Xvec_{lm};  \  \ 
 \Bvec^{(M)}=-\frac{i}{c}\nabla\crossprod
 \left[z_l(k_rr)\Xvec_{lm}\right].
\label{Mmultipole}
\ee
In both these equations the vector spherical harmonic 
\be
 \Xvec_{lm}=(\Lvec Y_{lm})/\sqrt{l(l+1)}.
\label{Xlmdef}
\ee
is defined in terms of the angular momentum operator 
$\Lvec\equiv -i{\bf r}\crossprod\nabla$ and
the usual spherical harmonics $Y_{lm}$.
The vector spherical harmonics fulfill the orthogonality relations
$
   \int d\Omega\, \Xvec_{l'm'}^*(\Omega)\cdot\Xvec_{lm}(\Omega)
   =\delta_{ll'}\delta_{mm'},
$
and
$
   \int d\Omega\, \Xvec_{l'm'}^*(\Omega)
   \cdot
   [\hat{r}\crossprod\Xvec_{lm}(\Omega)]
   =0
$
on the unit sphere.
The function $z_l(k_r r)$  stands for a
linear combination of a spherical Bessel function $j_l(k_r r)$, 
which is regular at $r=0$, and a spherical Hankel function  of the first kind
$h_l(k_r r)=j_l(k_r r) + i n_l(k_r r)$, which describes outgoing waves. 

In the present case, we can write the electric field inside the sphere 
at site $i$ as 
\be 
 \Evec=\sum _{lm} k\, c_{lm}^{(M)}\,  j_l(k_r r)\Xvec_{lm}+
 \frac{i}{\epsilon_r}\nabla\crossprod
 \left[c_{lm}^{(E)} j_l(k_rr)\Xvec_{lm}\right].
\ee 
Since the fields must be defined at the sphere center, only Bessel functions 
$j_l$ appear.
Just outside the sphere, we have instead
\barr 
 \Evec= &&\sum _{lm} 
  k\, [a_{lm}^{(M)}\,j_l(k r) + b_{lm}^{(M)} h_l(kr)]\, \Xvec_{lm}+
\breakeq
 + i\,\nabla\crossprod
 \left\{[a_{lm}^{(E)} j_l(kr) + b_{lm}^{(E)} h_l(kr)]\, \Xvec_{lm}\right\}.
\earr 

Demanding that the usual boundary conditions for the $\Evec$ and $\Bvec$
fields are satisfied at the surface of the sphere one can derive equations 
from which the field inside a sphere (the $c$ coefficient) and the
scattered field ($b$ coefficient) can be calculated for each multipole, given
the incident field ($a$ coefficient).  Since we focus on scattering 
properties, the key quantities are the ``sphere-response functions''
\be
  s_l^{(E)} = \frac{b_{lm}^{(E)}}{a_{lm}^{(E)}} 
=
 -\, \frac{\epsilon\, k R\, j_l'(k R) 
          + j_l (k R)\, (\epsilon-1-{\cal J}_l) }
   {\epsilon\, k R\, h_l'(k R) 
          + h_l (k R)\, (\epsilon -1 -{\cal J}_l) },
\label{slEres}
\ee
and
\be
 s_l^{(M)} = \frac{b_{lm}^{(M)}}{a_{lm}^{(M)}} =
 -
 \frac{ 
k R\, j_l'(k R) 
- 
j_l (k R)\, {\cal J}_l 
}
  { k R\, h_l'(k R) -  h_l (k R)\, {\cal J}_l },
\label{slMres}
\ee
where ${\cal J}_l$ is shorthand for
$$ {\cal J}_l= k_r R\, j_l'(k_rR)/j_l(k_rR). $$
Thanks to the symmetry of the sphere these 
response functions are independent of $m$.

In the following, we will often describe the ``state'' of a sphere through the
$a$ and $b$ coefficients.  It is therefore convenient to collect these
coefficients into vectors $\vec{a}_i$ and $\vec{b}_i$ in ``multipole
space'' with the structure
\be
  \vec{a}_i = \left( 
 a_{1-1,i}^{(E)};
 a_{10,i}^{(E)}; 
 \ldots;
 a_{l_{max}l_{max},i}^{(E)}; 
 a_{1-1,i}^{(M)};
 \ldots;
 a_{l_{max}l_{max},i}^{(M)} 
  \right),
\ee
etc.\cite{lmaxnote}
If we form a diagonal tensor $\tensor {s}$, 
in which the response functions $s_{lm}^{(E)}$ and
$s_{lm}^{(M)}$ enter in the appropriate places,
the relation between 
incoming and outgoing waves at site $i$ can be written
\be
 \vec{b}_i = \tensor{s} \vec{a}_i. 
\label{Sphereeq1}
\ee

\subsection{Surface response}

Light will be scattered also from the sample surface and we need to deal
with the surface response.
A plane wave incident on the sample surface  has a wave vector 
\be
   {\bf q}_{-} = {\bf q}_{\|} - \hat{z} \sqrt{k^2 - |{\bf q}_{\|}|^2}
\ee
(the subscript ``-'' indicates that the wave propagates in the negative $z$
direction)
and can be written 
\be 
  {\bf E}= \left\{ E^{(s)}  \hat{s}
 + E^{(p)} \hat{p} \right\}
  e^{i{\bf q}_{-}\cdot{\bf r}} 
\label{planewave1}
\ee
where the unit vectors for s and p polarization are 
$\hat{s}=[\hat{z}\crossprod \hat{q}_{\|}]$
and 
$\hat{p}=-[\hat{q}_{-} \crossprod (\hat{z}\crossprod\hat{q}_{\|})]$, 
respectively. Here 
$\hat{q}_{-}$ is a unit vector in the direction of the wave vector, and 
$\hat{q}_{\|}$ is a unit vector in the direction of
the in-plane component of the wave vector.

The reflected wave in response to the incident one (\ref{planewave1}) 
can be written
\be 
  {\bf E}_{\mathrm refl}= \left\{ \chi_s({\bf q}_{\|}, \omega) E^{(s)}  \hat{s}
 + \chi_p({\bf q}_{\|}, \omega) E^{(p)} \hat{p} \right\}
  e^{i{\bf q}_{+}\cdot{\bf r}},
\label{reflwave1}
\ee
where now 
${\bf q}_{+} = {\bf q}_{\|} + \hat{z} \sqrt{k^2 - |{\bf q}_{\|}|^2}$
and $\hat{p}=-[\hat{q}_{+} \crossprod (\hat{z}\crossprod\hat{q}_{\|})]$ 
for this wave vector.
In case the sample is {\em homogeneous} the surface response functions
$\chi_s$ and $\chi_p$ are found from the Fresnel formulae yielding
\be 
   \chi_s({\bf q}_{\|}, \omega)
 = \frac{\sqrt{k^2-{\bf q}_{\|}^2} - \sqrt{k_s^2-{\bf q}_{\|}^2} }
 {\sqrt{k^2-{\bf q}_{\|}^2} + \sqrt{k_s^2-{\bf q}_{\|}^2} },
   \label{chi_s}
\ee
where  $k_s$ is the wave number in the sample depending on the 
dielectric function of the sample $\epsilon_s(\omega)$,
$$
k_s^2=\epsilon_s(\omega)k^2 .
$$
For p polarized light one gets 
\be 
   \chi_p({\bf q}_{\|}, \omega)
 = \frac{\epsilon_s\sqrt{k^2-{\bf q}_{\|}^2} - \sqrt{k_s^2-{\bf q}_{\|}^2} }
 {\epsilon_s\sqrt{k^2-{\bf q}_{\|}^2} + \sqrt{k_s^2-{\bf q}_{\|}^2} }.
   \label{chi_p}
\ee

Note that
the expressions for all the surface response functions can be extended to
the case of evanescent waves for which $|{\bf q}_{\|}| > k$ (when
evaluating square roots the branch cut lies below the positive real axis).
For lack of a better terminology, ``plane wave'' will
sometimes be used to describe both propagating plane waves and waves that
propagate in the directions parallel to the sample surface but are
evanescent in the third, $z$, direction.

When the sample consists of several layers of different materials the
response functions $\chi_p$ and $\chi_s$ must be calculated by a 
generalized approach, for example a transfer matrix formalism, see Ref.\ 
\onlinecite{Zak}. Then one first makes an Ansatz for the electromagnetic
field in the bottom layer of the sample in terms of one plane wave that
propagates or decays exponentially in the downward direction. Then this wave
is matched to a downgoing and an upgoing wave at the bottom interface. These 
waves are propagated through the next layer, matched at the next interface,
etc., until one reaches the sample surface where the ratio between the
incident and upgoing wave amplitudes yields the surface response for the
two different polarization types.

\subsection{Multiple-scattering solution for an ordered array}

Having introduced the necessary basic ingredients of the calculation
we consider now the complete array of spheres (without the sample for the
moment), and  
let a plane wave like the one  in Eq.\ (\ref{planewave1})
impinge on this system.
Now Eq.\ (\ref{Sphereeq1}) can be written
\be
 \vec{b}_i = \tensor{s} \left[\vec{a}_i^{\rm dir} 
    + \sum_{j\neq i} \tensor{t}_{ij}^{\rm dir} \vec{b}_j \right]
\label{Sphereeq2}.
\ee
Thus the incident field at sphere $i$ is,
as can be understood by looking at Eqs.\ (\ref{KirchE}) and 
(\ref{KirchB}), a sum of one contribution,
$\vec{a}_i^{\rm dir}$, 
coming  from the projection of the plane wave 
in Eq.\ (\ref{planewave1})
onto the various multipoles,
and contributions from waves scattered off all the other spheres.
The explicit, rather lengthy, expressions for $\vec{a}_i^{\rm dir}$ 
and the coupling coefficients $t_{ij}^{\rm dir}$ (or rather its Fourier
transform) are derived in Appendices
\ref{Appendplane} and \ref{Appendtdir}, respectively.

Thanks to the periodicity of the sphere array, $\tensor{t}_{ij}$ only
depends on the relative vector ${\bf R}_i - {\bf R}_j$ separating
spheres $i$ and $j$, and 
Eq.\ (\ref{Sphereeq2}) can be solved  by a Fourier transformation.
For an arbitrary quantity $Z$ defined on the lattice we introduce the
Fourier transform 
\be 
   Z_{\bf Q} = \sum_{i} Z_i e^{-i{\bf Q}\cdot {\bf R}_i},
\label{Fourier}
\ee
and its inverse 
\be 
   Z_i = \frac{1}{N} \sum_{\bf Q} Z_{\bf Q} e^{i{\bf Q}\cdot {\bf R}_i},
\label{Fourierinv}
\ee
where $N$ is the number of lattice sites.
Applied to Eq.\ (\ref{Sphereeq2}) this yields
\be
  \vec{b}_{\bf Q} = \tensor{s} \left[
  \vec{a}_{\bf Q}^{\rm dir} + \tensor{t}_{\bf Q}^{\rm dir} \vec{b}_{\bf Q}
\right], 
\ee 
so that
\be
 \vec{b}_{\bf Q} = \left[
\tensor{1}-\tensor{s} \tensor{t}_{\bf Q}^{\rm dir}\right]^{-1} 
   \tensor{s}  \vec{a}_{\bf Q}^{\rm dir}.
\ee

When the sample is added to the problem one more surface, the plane 
$z = z_0=-R$, will contribute to the Kirchoff integrals. The waves that
are sent out from the sample surface are in turn emitted in response 
to the incident plane wave or waves coming from the spheres.
Consequently Eq.\ (\ref{Sphereeq1}) now reads
\be
 \vec{b}_i = \tensor{s} \left[\vec{a}_i^{\rm dir} + \vec{a}_i^{\rm ref}
    + \tensor{w} \vec{b}_i
 + \sum_{j\neq i} (\tensor{t}_{ij}^{\rm dir} + \tensor{t}_{ij}^{\rm sub})
       \, \vec{b}_j
 \right],
\label{Sphereeq3}
\ee
where $\vec{a}_i^{\rm ref}$ is the contribution from the incident wave after
it has been reflected {\em once} from the sample, $\tensor{w}$ is the 
sample-mediated self-interaction between the various multipoles on a sphere,
while $\tensor{t}^{\rm sub}$ is the sample-mediated interaction 
between two different spheres. These quantities are discussed in Appendix 
\ref{Appendtsub}.
It should be kept in mind that both $\tensor{w}$ and $\tensor{t}^{sub}$
describe events in which the source sphere sends out a wave that is 
scattered once off the sample and then goes directly to the receiving sphere;
multiple scattering events enter through Eq.\ (\ref{Sphereeq3}).
We introduce a total driving field
\be 
 \vec{a}_i^{\rm ext} = \vec{a}_i^{\rm dir} + \vec{a}_i^{\rm ref},
\ee
and a total sphere-sphere interaction
\be 
  \tensor{t}_{ij}= \tensor{t}_{ij}^{\rm dir} + \tensor{t}_{ij}^{\rm sub},
\ee
and Fourier-transforming we find the solution for $\vec{b}_{\bf Q}$
\be
 \vec{b}_{\bf Q} = \left[\tensor{1} - \tensor{s} \tensor{w} 
   - \tensor{s} \tensor{t}_{\bf Q}\right]^{-1} 
   \tensor{s}  \vec{a}_{\bf Q}^{\rm ext}.
\ee

It is also possible to rewrite Eq.\ (\ref{Sphereeq3}) as 
\be 
   \vec{b}_i= (\tensor{1}-\tensor{s}\tensor{w})^{-1}\tensor{s}
   \left[\vec{a}_i^{\rm ext} 
    + \sum_{j\neq i} \tensor{t}_{ij} 
       \, \vec{b}_j
   \right].
\label{Sphereeq4}
\ee
Thus, the sample-mediated self-interaction does not appear explicitly in
this equation, but is instead accounted for by the modified response
tensor $(\tensor{1}-\tensor{s}\tensor{w})^{-1}\tensor{s}$.  Interparticle
interactions mediated by the sample are of course still included in
$\tensor{t}_{ij}$.  The reasoning behind Eq.\ (\ref{Sphereeq4}) in many
ways makes it easier to think of the problem, since the sample in a
sense has been eliminated from the formalism. The effects of the sample on
the spheres are summarized by the modification of the response tensor and
the additional contribution $\tensor{t}_{ij}^{\rm sub}$ to the
particle-particle interaction.

\section{Disordered array}
\label{SecDisorder}

\subsection{Coherent potential approximation}
\label{CPAsec}

In the previous section we dealt with a system where each site on the
lattice was occupied by a sphere.  We will now treat a disorder model,
namely a lattice gas. We still have a square lattice but only a random
fraction $p$ of the sites are occupied by spheres, whereas the rest of the
sites are empty.

The averaging over different disorder realizations will be done within the
coherent-potential approximation.\cite{Rick}  The CPA is known to be the best
single-site mean-field theory, and it gives correct results in a number of
important limits.  A certain realization of the disordered system is
characterized by the response function of the sphere occupying the various
sites; at the empty sites this response ($\tensor{s}$) vanishes.  Within
the CPA calculation we aim at finding an effective medium in which {\em
every} site is occupied by average objects that can also be described by a
surface response tensor that we will denote $\tensor{\alpha}$ in the
following. As we saw above, the treatment can be simplified by
incorporating the sample-mediated self interaction into the particle
response. For the average particles we do this by also introducing the tensor
$\tensor{\beta}$ which is related to $\tensor{\alpha}$ by
\be
   \tensor{\beta}= 
   \left(\tensor{1}-\tensor{\alpha}\tensor{w}\right)^{-1}
   \tensor{\alpha}, 
   \ \ \Leftrightarrow \ \ 
   \tensor{\alpha}= 
   \left(\tensor{1}+\tensor{\beta}\tensor{w}\right)^{-1}
   \tensor{\beta}. 
\ee

It should already now be pointed out that the tensors $\tensor{\alpha}$ and
$\tensor{\beta}$, describing the average scatterers, are not diagonal; the
spherical symmetry is lost when interactions with the environment are taken
into account.
For example,
in the case that the particles are treated as dipoles, the average
particles have different polarizabilities in the directions parallel to,
and normal to the plane of the array, respectively.\cite{PerLi}
Both $\tensor{\alpha}$ and $\tensor{\beta}$ are, however, 
still lattice symmetric. Thus, only those tensor elements that couple
multipoles belonging to the same irreducible representation of the
array point group (in the present case where the particles interact 
with a sample, $C_{4v}$) are nonzero.

The response functions $\tensor{\alpha}$  and $\tensor{\beta}$ are determined
self-consistently by placing a real site, i.e.\ one which is occupied by a
real sphere with probability $p$ and empty with probability $1-p$ in the
effective medium of average particles, and
demanding that the real site does not cause any  additional scattering on
the average.

To see what this means in detail, we consider the scattering off the real
site as compared with when it is occupied by an average scatterer.  An
occupied real site yields the {\em extra} outgoing waves
\be 
   \vec{b}_{\rm 1, extra} = (\tensor{\beta}_{\rm oc}-\tensor{\beta})
   \, \vec{a}_0
\label{bextra1}
\ee
in a {\em single scattering event.} In Eq.\ (\ref{bextra1})
\be 
   \tensor{\beta}_{\rm oc}=
   \left(\tensor{1}-\tensor{s}\tensor{w}\right)^{-1}\tensor{s}, 
\ee
the effective response function for a real sphere including the
sample-mediated self interaction,
and 
$\vec{a}_{0}$ could describe any combination of waves incident on the real
site. In the CPA one includes not only single scattering events off the
real site, but takes into account multiple scattering to all orders. This
means that the extra scattering from an occupied site can be written with
the aid of a $T$ matrix $\tensor{T}_{\rm oc}$,
\be
   \vec{b}_{\rm oc, extra} = 
   \tensor{T}_{\rm oc} \,
   \vec{a}_0
   =
   \left[\tensor{1}-(\tensor{\beta}_{oc}-\tensor{\beta})
   \tensor{\gamma}_{00}\right]^{-1} 
   (\tensor{\beta}_{oc}-\tensor{\beta})
   \vec{a}_0.
\label{T_oc}
\ee
Here $\tensor{\gamma}_{00}$ is a propagator for multipole
excitations describing how waves sent out from the real site propagate
through the effective medium of average scatterers and eventually return to
the real, central site. One can show that
\be
    \tensor{\gamma}_{00} =
  N^{-1} \sum_{\bf Q} 
   \left(\tensor{1}-\tensor{t}_{\bf Q} \tensor{\beta}\right)^{-1} 
   \tensor{t}_{\bf Q}.
\label{gamma0def}
\ee
If instead the real site is empty, the extra scattering can be written
\be
   \vec{b}_{\rm un, extra} = 
   \tensor{T}_{\rm un} \,
   \vec{a}_0
   =
   \left[\tensor{1}-(\tensor{\beta}_{un}-\tensor{\beta})
   \tensor{\gamma}_{00} \right]^{-1} 
   (\tensor{\beta}_{un}-\tensor{\beta})
   \vec{a}_0.
\label{T_un}
\ee
Of course in this case the empty site does not give any scattering, 
i.e.\ $\tensor{\beta}_{un}=\tensor{0}$, and
\be
   \vec{b}_{\rm un, extra} = 
   \left[\tensor{1}+\tensor{\beta}
   \tensor{\gamma}_{00}\right]^{-1} 
   (-\tensor{\beta})
   \vec{a}_0.
\ee

The CPA self-consistency condition demanding that there should be {\em no
extra scattering on the average} means that
\be
   p \, \vec{b}_{\rm oc, extra} +
   (1-p) \, \vec{b}_{\rm un, extra} =0.
\label{CPA1}
\ee
After some algebra Eq.\ (\ref{CPA1}) can be rewritten as 
\be 
   \tensor{\beta}=
   p \, \tensor{\beta}_{oc}
   \left\{
      \tensor{1} + 
      \tensor{\gamma}_{00} \, 
      \left[\tensor{\beta}-\tensor{\beta}_{oc}\right]
   \right\}^{-1}.
\label{CPA2}
\ee
Equation (\ref{CPA2}) has to be solved for $\tensor{\beta}$ by an 
iterative process in
which also $\tensor{\gamma}_{00}$ in Eq.\ (\ref{gamma0def}) has to be
updated in each step of the iteration.

From  Eq.\ (\ref{CPA2}) it is easy to verify one of the celebrated features
of the CPA, namely that it approaches the correct limit for both $p=0$ when
$\tensor{\beta}=0$, and $p=1$ for which the self-consistent solution is
$\tensor{\beta} =\tensor{\beta}_{oc} =
(\tensor{1}-\tensor{s}\tensor{w})^{-1}\tensor{s}$.

\subsection{Diffuse scattering}
\label{Diffuscattsec}

So far we have found the response functions $\tensor{\alpha}$ and
$\tensor{\beta}$ that characterize an average effective medium that has the
same effect as the lattice gas of particles on the incident light in terms
of {\em coherent} transmission and reflection.

To calculate the intensity of diffusely scattered light into a certain
direction, the theory has to be developed further. Up to now we have only
calculated average quantities involving one single multipole excitation,
but to find diffuse scattering {\em intensities} averages involving two 
multipole excitations must be calculated.

Suppose that a plane wave described by the amplitude vector $\vec{a}_{\bf
Q}^{\rm ext}$ impinges on the disordered particle array.  In response to
this there will of course be multipole excitations  at the incident wave
vector ${\bf Q}$ that can be characterized by $\vec{b}_{\bf Q}$. 
But as a result of the disorder there are also multipole excitations
described by $\vec{b}_{\bf Q'}$ at other wave vectors ${\bf Q'}$. 
Now, averaging over a large number of different disorder realizations (in
our case, lattice gas arrangements) would show that $\langle \vec{b}_{\bf
Q} \rangle \ne 0$, (here $\langle X \rangle$ denotes disorder-averaging of
the quantity $X$) whereas $\langle \vec{b}_{\bf Q'} \rangle = 0$, since the
diffusely scattered light has an essentially random phase that varies from
one disorder realization to another. Yet averages like 
$\langle b_{\xi,{\bf Q'}}^* b_{\zeta,{\bf Q'}} \rangle$ ($\xi$ and $\zeta$ 
are shorthand indices for different multipoles)  of course do {\em not}
vanish and hence there is diffusely scattered light.

Assume that we have managed to calculate the coefficients
$\vec{b}_{\zeta{\bf Q'}}$ for one particular realization of the disorder.
The (reflected) scattered electric field  $ {\bf E}_{sc} ({\bf r})$,
with a wave vector
$$
   {\bf q'}={\bf q'}_{\|} + q_z' \hat{z}
   {\rm \ \ with  \ \ }
   q_z' = \sqrt{k^2-q_{\|}'^2} 
$$
whose in-plane wave vector  
${\bf q'}_{\|} ={\bf Q'}+ {\bf G}$ equals ${\bf Q'}$ up to a reciprocal
lattice vector ${\bf G}=(2\pi/a) (n_x \hat{x}+n_y\hat{y})$ with $n_x$ and
$n_y$ being integers,
can then be written
\be
   {\bf E}_{sc}({\bf r}) 
   =
   \int \frac{d^2q'_{\|}}{(2\pi)^2}
   \sum_{\zeta}
   b_{\zeta{\bf Q'}}
   \left[
      \hat{s} d_{\zeta}^s({\bf q'})
      +
      \hat{p} d_{\zeta}^p({\bf q'})
   \right].
\ee
Here the function
\be 
   d_{\zeta}^s({\bf q'})
   =
   g_{\zeta}^s({\bf q'}_{+})  
   +
   e^{2i\sqrt{k^2-q_{\|}'^2}|z_0|}      
   \chi_s({\bf q'}_{\|}) g_{\zeta}^s({\bf q'}_{-}) 
\ee
describes coupling from multipole $\zeta$ to an outgoing s polarized wave,
either directly (first term) or through a reflection off the sample (second
term). The functions $g_{\zeta}^{s,p}$  which are defined in Appendix
\ref{Appendplane}, measure the ``overlap'' between multipole $\zeta$
and a plane wave with wave vector 
${\bf q'}_{\pm} = {\bf q'}_{\|} \pm q_z' \hat{z}.$ 
When the wave leaves the system without being reflected off the
sample the wave vector should have a positive $z$ component, whereas
when a reflection takes place the multipole must first couple to a
down-going plane wave (${\bf q'}_{-}$), and the amplitude of the wave is
also affected by a phase shift due to the additional distance that the wave
must travel and the surface response function $\chi_s$.  In the same way,
\be 
   d_{\zeta}^p({\bf q'})
   =
   g_{\zeta}^p({\bf q'}_{+})  
   +
   e^{2i\sqrt{k^2-q_{\|}'^2}|z_0|}      
   \chi_p({\bf q'}_{\|}) g_{\zeta}^p({\bf q'}_{-}) 
\ee
for outgoing p polarized waves.
The scattered power is now for this particular realization
\barr 
   P_{sc} && = \frac{c \, \epsilon_0}{2}
   \int \frac{d^2q'_{\|}}{(2\pi)^2}
   \sum_{\xi\zeta}
   b_{\xi{\bf Q'}}^* b_{\zeta{\bf Q'}}
\breakeq
\times
   \left[
      d_{\xi}^s({\bf q'})^* d_{\zeta}^s({\bf q'})
      +
      d_{\xi}^p({\bf q'})^* d_{\zeta}^p({\bf q'})
   \right].
\label{power_realize}
\earr

If we were to calculate the scattered power for {\em another}
realization of the disorder we would arrive at the same formal expression 
(\ref{power_realize}); the functions $d_{\xi}^{s(p)}({\bf q'})$ would still
be the same, however, the values of the $b$ coefficients would 
of course be different.
Thus, the disorder-averaged scattered power is given by Eq.\
(\ref{power_realize}) provided that the product 
$ b_{\xi{\bf Q'}}^* b_{\zeta{\bf Q'}} $ is replaced by its disorder-average
$$
   b_{\xi{\bf Q'}}^* b_{\zeta{\bf Q'}}  \rightarrow
   \left\langle b_{\xi{\bf Q'}}^* b_{\zeta{\bf Q'}} \right\rangle.
$$
Obviously, to find the intensity of diffusely scattered light we must 
calculate the averages
$\langle b_{\xi{\bf Q'}}^* b_{\zeta{\bf Q'}} \rangle$.

\subsection{Vertex function}

Since we have calculated the properties (essentially $\tensor{\beta}$) of 
the average objects within the CPA, also the calculation of 
$\langle b_{\xi{\bf Q'}}^* b_{\zeta{\bf Q'}} \rangle$
must be done at the same
level of approximation in order to conserve the total energy. 
This means that this calculation must take into account repeated
single-site scattering events to all orders.  To this end  we have to solve
a Bethe-Salpeter equation for a vertex function $\Lambda$ that sums all 
``ladder diagrams'' as illustrated in Fig.\ \ref{Bethefig}.\cite{Mahan} 

The irreducible vertex $\Xi$ describes the simplest scattering event, off
one real site, that enters the sum of ladder diagrams. It is given by an
average between the scattering intensity from an occupied site and from an
unoccupied site, and can be expressed in terms of the previously introduced
$T$ matrices [see Eqs.\ (\ref{T_oc}) and (\ref{T_un})] as
\be
    \Xi_{\zeta\delta}^{\xi\eta} =
    p\, 
     (\tensor{T}_{\rm oc}^*)_{\xi\eta}
     (\tensor{T}_{\rm oc})_{\zeta\delta}
   +
    (1-p) 
     (\tensor{T}_{\rm un}^*)_{\xi\eta}
     (\tensor{T}_{\rm un})_{\zeta\delta}.
\ee

The full vertex function $\Lambda$ is then found by allowing for the
possibility of consecutive scattering events off different sites. Summed to
all orders this yields the Bethe-Salpeter equation (summation over repeated
indices is assumed implicitly)
\be
   \Lambda_{\zeta\delta}^{\xi\eta}
   =
   \Xi_{\zeta\delta}^{\xi\eta}
   +
   \Xi_{\zeta\delta'}^{\xi\eta'} \,
   \Pi_{\delta'\delta''}^{\eta'\eta''} \,
   \Lambda_{\delta''\delta}^{\eta''\eta}.
\label{BetheS}
\ee
In Eq.\ (\ref{BetheS}), $\Pi$ describes the propagation of the multipole
excitations through the effective medium from one scattering site to the
next.  This propagator involves essentially the same quantities as did the
``return'' propagator $\tensor{\gamma}_{00}$ encountered earlier. We can 
write 
\be
   \Pi_{\zeta\delta}^{\xi\eta} 
   =
   -
   (\tensor{\gamma}_{00})_{\xi\eta}^*  (\tensor{\gamma}_{00})_{\zeta\delta}
   +
   N^{-1} \sum_{\bf Q} 
   (\tensor{\gamma}_{\bf Q}^*)_{\xi\eta}
   (\tensor{\gamma}_{\bf Q})_{\zeta\delta},
\ee
where
\be 
   \tensor{\gamma}_{\bf Q}
   =
   \left(\tensor{1}-\tensor{t}_{\bf Q} \tensor{\beta}\right)^{-1} 
   \tensor{t}_{\bf Q}.
\label{gammaQdef}
\ee
Inserting the above expressions for $\Xi$ and $\Pi$ into Eq.\
(\ref{BetheS}) we can solve for $\Lambda$ numerically by means of a matrix
inversion.

Finally, we must relate the vertex function to the averages 
$\langle b_{\xi{\bf Q'}}^* b_{\zeta{\bf Q'}} \rangle$
needed to find the diffuse scattering intensity.  One can write
\barr
   \left \langle b_{{\bf Q'}\xi}^*b_{{\bf Q'}\zeta} \right \rangle 
   &&
   =
   \left[ 
      (\tensor{1} - \tensor{\beta}\tensor{t}_{\bf Q'})^{-1}
   \right]_{\xi\xi'}^*
   \left[ 
      (\tensor{1} - \tensor{\beta}\tensor{t}_{\bf Q'})^{-1}
   \right]_{\zeta\zeta'}
\breakeq
\times
   \Lambda_{\zeta'\delta}^{\xi'\eta} \, \, \,
   (\vec{a}^{{\rm eff}*})_{\eta} \,
   (\vec{a}^{\rm eff})_{\delta}.
\label{av_from_vert}
\earr
Here 
\be
   \vec{a}^{\rm eff} 
   =
   N^{-1} \sum_{\bf Q}
   \left(\tensor{1} - \tensor{t}_{\bf Q} \tensor{\beta} \right)^{-1}
   \vec{a}_{\bf Q}^{\rm ext}
\label{aeffdef}
\ee
represents the effective driving field at the site where the first scattering
event occurs.  This effective field results from screening of the external
field (given by $\vec{a}_{\bf Q}^{\rm ext}$) by the average particles in
the effective medium.   In the same way, the tensor 
$(\tensor{1} - \tensor{\beta}\tensor{t}_{\bf Q'})^{-1}$
appearing in front of $\Lambda$ in Eq.\ (\ref{av_from_vert}) accounts for
the screening of the outgoing waves by the average particles after the last
scattering event has occurred. One should note that the outgoing wave
vector ${\bf Q'}$ occurs in this expression, whereas, in the case of an
incident plane wave, there is only one term (at wave vector ${\bf Q}$) in
the sum defining $\vec{a}^{\rm eff}$.

\section{Results and discussion}
\label{SecRes}

Let us consider a system of spheres with radius $R=50$ nm and 
lattice parameter $a=120$ nm.
These values probably describe the average properties of the
particle overlayers studied by Stuart and Hall\cite{SH} rather well.
However, one should keep in mind that in that experiment the 
typical particle shape was most likely not spherical. 
Judging from the manufacturing process, the particles
studied in that experiment looked more like oblate spheroids than spheres.

Figure \ref{lambdafig} shows calculated results for the intensity of
scattered light from a disordered array of silver particles placed on a  
Si/SiO$_2$/Si sample.
In close agreement with the experimental results found in Ref.\
\onlinecite{SH} the scattering intensity in Fig.\ \ref{lambdafig} (a)
has two isolated peaks at 1180 and
710 nm.\cite{expres} 
These isolated peaks are then followed by a broader ``scattering
band'' at shorter wavelengths $\alt 550 \unit{nm}$.
We also note that the overall intensity of the scattered light 
is in reasonable
agreement with the experimental results; the measured diffuse scattering
intensity collected over a solid angle of $\approx 2\pi/5$ was about 1 \%
of the incident intensity, while we here get intensities of the order of 10
\% when calculating the light collected over a solid angle of $2\pi$.

What mechanism makes the two isolated, long-wavelength peaks appear in the
spectrum?  These features only appear when there is a layered sample; as can
also be seen in Fig.\ \ref{lambdafig} (a) there are no such peaks in the 
spectrum calculated with a homogeneous dielectric sample (LiF).
Stuart and Hall \cite{SH} argued that  the intensity of the scattered light
reaches a maximum whenever the particle-particle interaction 
mediated by waveguide modes in the sample reaches a maximum.
While this certainly is an interesting suggestions the results of our
calculation show that the isolated resonances have a simpler explanation.

The isolated peaks in the scattering intensity coincide with maxima
in the electric field that polarizes the metallic particles in the first
place.  The dotted curve in Fig.\ \ref{lambdafig}(a)  shows the 
squared strength of
the driving electric field  found in the plane of the sphere centers as a
result of the incident wave and its reflection off the sample.  This field
has been calculated in the absence of the particle overlayer. In the
present situation, when the incident field hits the sample at normal
incidence (no physical difference between s and p
polarization), the driving field can be written
\be
   {\bf E}_{\mathrm drive}=  E^{(s)}  \hat{s}
   [1+e^{2 ik|z_0|}  \chi_s({\bf q}_{\|}=0, \omega)],
\label{drive1}
\ee
thus the quantity plotted together with the scattering intensity in Fig.\
\ref{lambdafig} is 
$$
   \left| 1+e^{2 ik|z_0|}  \chi_s({\bf q}_{\|}=0, \omega) \right|^2.
$$
The
oscillations in the driving field strength with varying wavelength is 
consequently a result of
constructive and destructive interference between waves reflected from
different interfaces of the multilayer sample.
For the case shown in Fig.\ \ref{lambdafig} (a), 
even if  the presence of the other layers also affects the 
driving field, the thickness $d_1$, of the top Si layer is the most
important parameter in determining the interference maxima and minima;
$\lambda_{\mathrm Si}=\lambda/\sqrt{\epsilon_{\mathrm Si}}$ equals $2d_1$
when $\hbar\omega\approx$ 1.1 $\unit{eV}$ ($\lambda\approx$1130 nm), and $4d_1$
when $\hbar\omega\approx$ 1.9 $\unit{eV}$ ($\lambda\approx$650 nm).
Of course, when one changes the sample geometry i.e.\ the thicknesses 
or the composition of the various layers, the maxima of the driving field
shift in frequency.  As can be seen in Fig.\ \ref{lambdafig} (b), where the
thickness $d_1$ of the top Si layer was increased to 240 $\unit{nm}$, this
brings with it identical shifts of the isolated peaks in the scattered
intensity.  This further supports the interpretation given here.

Figure \ref{fieldfig} shows how the in-plane electric field varies with $z$
in the Si/SiO/Si sample used for the calculations in Fig.\ \ref{lambdafig}
(a) when there is no particle overlayer and waves with photon energies 
1.1 eV, 1.3 eV, 1.5 eV, and 1.7 eV,
respectively, impinge on the surface at normal incidence. 
As can be seen there are reflection resonances when it is possible to fit
in a wave that has antinodes at the interfaces of the top Si layer.

A simple explanation for why surface waveguide modes are {\em not} causing
the scattering resonances lies in the very small scattering
cross section for a single sphere at the relevant wavelengths.   
This cross section is of the
order $(kR)^4 \, R^2$.  With a wavelength of $1000\, \unit{nm}$ and $R=50\,
\unit{nm}$ $kR \sim 0.3$, which means that the scattering cross section is
of the order of 1 \% of the geometric cross section. 
Now, the process
discussed by Stuart and Hall\cite{SH}
involves {\em two} scattering events off nanoparticles 
because the incoming light must first be scattered into a waveguide mode by
the particles and then out again in order to be observed as
diffusely scattered light.
Of course  such a process will not be very
effective, unless some other factor compensates for the small scattering
cross sections.  In this case the waveguide modes do offer some
compensation; the response of the semiconductor surface is very strongly
enhanced for combinations of in-plane wave vector ${\bf q}_{\|}$ and photon
energy $\hbar\omega$ that coincide with those of a waveguide mode. But
this only happens for a narrow interval of wave vectors. 
There is not enough phase space for the waveguide-mode-mediated
particle-particle interaction to become an important factor for the
scattering of long-wavelength light.

This point is best illustrated by looking at a plot of the particle-particle
interaction as a function of in-plane wave vector shown in 
Fig.\ \ref{couplfig}.
The curves show the value of $|\beta_{11,11{\bf Q}}t_{11,11{\bf Q}}|$
as a function of ${\bf Q}$ for two different photon energies.  
This quantity yields the ratio 
between the secondary wave hitting a sphere as a result of a previous
scattering event off the other average particles, and the primary incident
wave. With this particular choice of tensor elements we are focusing the
attention to the waves sent out by in-plane electric dipoles on the
particles.
We note that for most wave vectors the
sphere-sphere coupling for these photon energies lie well below 1, i.e.\
the primary wave dominates in strength over rescattered secondary waves.
For certain wave vectors the sphere-sphere coupling is, however, rather
strong. This happens when the electromagnetic oscillations of a particle
are in resonance with a waveguide mode of the sample. In this case the
sample-mediated coupling has a sharp peak. With 
$\hbar \omega=1.0\,\unit{eV}$ there are two such resonances. The peak
marked A$'$ is due to an s polarized waveguide mode whereas the peak B$'$ is
caused by a p polarized mode. When increasing the photon energy these peaks
move out towards higher ${\bf Q}$. 
At 1.3 eV their corresponding positions
are given by the peaks A and B. The new peak C 
is associated with a standing wave resonance very
similar to the one occurring at ${\bf Q}=0$ at 1 eV.

In this context we should also say that the waveguide modes may play a much
more pronounced role in experiments with periodically ordered arrays of
nanoparticles. In this case the surface modes that can be excited must have
an in-plane wave vector that either equals the incident one or else differs
from it by a reciprocal wave vector ${\bf G}$. 
In that case the sample-mediated particle-particle interaction is not
averaged over different wave vectors in the same way as in the present
case.  The recent experiment
by Linden {\it et al}.\cite{Lind} who measured light transmission through a
layered sample covered by an array of gold nanoparticles, provides such an
example.

Turning to the range of the spectrum at higher photon energies,
the features appearing there 
are caused by a combination of particle-sample and 
particle-particle interactions.
The disorder in the sphere array of course leads to considerable
broadening of the spectrum.
To analyze this in more detail, we display in Fig.\ \ref{coveragefig}
(using a photon-energy scale)
the diffuse scattering intensity for three different coverages 
$p=$0.1, $p=$0.2, and $p=$0.5, respectively.
It is evident that the particles do not act independently of each other
already from the fact that the scattering intensity as a function of
coverage saturates for photon energies above 2 eV or so. 
Overall, $p=0.2$ gives a stronger scattering than $p=0.5$.
This is in contrast with the situation at the peaks at 1.1 eV and 1.7 eV
where the scattering intensity is basically proportional to $p$, consistent
with a picture where the different particles scatter light essentially
independently of each other.

While the effects of particle-particle interactions are less clean-cut in
the previous case than when the particles are much smaller than 
$\lambda/2\pi$ (cf.\ Ref.\ \onlinecite{PerLi}),
we can still gain some more insight into what is happening in the various 
parts
of the ``scattering band'' by looking at the field distribution around the
average objects at different photon energies.\cite{lmaxnote} 
Figure \ref{vecfig} shows the
electric field around an average scatterer at 2.5 eV, 2.9 eV and 3.6 eV,
when an array of such scatterers are placed on a Si/SiO$_2$/Si sample with 
$d_1 =160\,\unit{nm} $ and $d_2=200\, \unit{nm}$ and illuminated by light
at normal incidence. The solid arrows represent the real part of the field
(in phase with the incident wave  in the plane  $z=0$).
(i) At  2.5 eV there is a concentration of the field to the region in
between the particles.
This is a result of direct 
particle-particle interactions
where a particle is polarized by the effective field which is the sum of the 
external field and the field from the other particles, the nearest neighbors 
giving the largest contributions.  
Indeed, if one considers two (real) silver spheres with $R=50\,\unit{nm}$
at a distance 120 nm from each other that are driven by an 
external field  directed along
the line joining  the two spheres, they have a collective resonance  at
about 2.6 eV. Thus this resonance is 
redshifted relative to the dipole resonance of an isolated particle which
occurs at about 3.1 eV if $R=50 \unit{nm}$  since the field from the
neighouring sphere acts in phase with the external field when the frequency
lies below the single-sphere resonance frequency.
In Fig.\ \ref{coveragefig} we see that as the coverage is increased, the
low-frequency edge of the scattering band  shifts to the left and the
intensity in this part of the spectrum increases, consistent with the
growing importance of particle-particle interactions at a larger $p$. 
(ii) Looking at the fields at  2.9 eV we see that they are more
concentrated to the region in space near the sample. 
Here particle-sample
interactions play a relatively important role.
What happens is that the external field polarizes the 
particles, and the particle dipole moments, in turn, induce 
image charges in the sample that reinforce the driving electric field.
In this way the dipole, but also higher multipole moments on the particle
are further amplified, leading to the field concentration near the sample
surface.
(iii) 
At 3.6 eV, on the other hand, the fields are concentrated to the upper
(vacuum) side of the particles, and the fields are also weaker in this case
than in the previous ones.  Here the induced dipole moment on a sphere is
essentially out of phase with the driving field.  Hence the image charges
in the sample this time set up an induced field opposing the external
electric field acting on the particle.   The enhancement mechanisms
discussed above are no longer efficient and the fields are localized to the
top side of the particle.

In their experiments, Stuart and Hall\cite{SH} also studied light
scattering from an overlayer of Ag particles on a silver sample covered by
spacer layers of LiF of varying thickness. The silver surface possesses
elementary excitations in the form of surface plasmons and 
these excitations can mediate interactions
between the silver particles in the overlayer.  
Also in this case, the experimental results exhibited scattering
resonances, albeit very broad ones, at relatively small photon energies.
In the experimental paper\cite{SH} they were discussed along the same 
lines as the
resonances ocurring in the case of the layered Si/SiO$_2$/Si sample, namely as
the result of strong surface-mediated particle-particle interactions that
enhance the scattering intensities.

Figure \ref{silverfig} shows results for the  calculated scattering 
intensity from a
particle overlayer on silver samples with LiF spacer layers of two
different thicknesses, 120 nm and 200 nm.
With a Ag sample the calculations are numerically more difficult than with
a Si sample because the surface response now, due to surface plasmons, has
considerably sharper features than what one sees in Fig.\ \ref{couplfig}.
There is therefore some numerical noise in the results, as can be seen as
irregular oscillations of the curves in Fig.\ \ref{silverfig}. 

The results in Fig.\ \ref{silverfig} agree well with the 
experimental ones from a qualitative point of view.
The shape of the spectra are the same as in the experiment, even if the
positions  of the maxima are somewhat shifted compared with the results in
Fig.\ 4 of Ref.\ \onlinecite{SH}. 
Part of the explanation for this may be that the height of the particles is
smaller in the experiment than in the calculation.
As was the case
with a Si/SiO$_2$/Si sample, the calculated  
intensity of the scattered light closely 
follows the
strength of the driving field, which is also plotted in the figure.
We can conclude that also for the silver sample the overall shape
of the diffuse scattering spectrum is set by the driving electric field
which in turn is affected by interference effects between waves reflected
from the different interfaces in the sample.
Increasing the thickness of the LiF layer from 120 nm to 200 nm shifts
both the peaks in the scattering spectrum towards longer wavelengths. This
brings with it a decrease in the scattering intensity at long wavelengths
since the maximum of the driving field falls at longer wavelengths where
the scattering cross section of the particles is smaller.  The scattering
intensity at the short-wavelength peak, on the other hand, increases when 
$d$ increases from 120 nm to 200 nm.  
Both these trends for the scattering intensity are qualitatively consistent
with the experimental results. 
The strength of the short-wavelength peak grows when $d$ changes from 120
nm to 200 nm because the maximum of the driving field in the latter case
coincides with the previously discussed resonance near 2.5 eV at which the
in-plane dipole moments of neighbouring particles oscillate in phase [cf.\
Fig.\ \ref{vecfig} (a)].

\section{Summary}
\label{SecSummary}

In this paper we have presented a formalism by which the diffuse light 
scattering intensity from a disordered array of spherical nanoparticles can
be calculated. The theory is built on an expansion of the electromagnetic
field around the nanoparticles in terms of multipoles, and takes into
account interparticle interactions mediated by near and far fields as well
as the sample.   The theory was then applied to systems similar to those
studied experimentally by Stuart and Hall\cite{SH} in which silver 
nanoparticles are placed on layered substrates that support various surface
waves.  The results of the present calculation agrees rather well with the
experimental ones. In particular the intensity of diffusely scattered light
from nanoparticles on a layered Si/SiO$_2$/Si sample shows several
characteristic peaks at long wavelengths. We have shown that these
peaks are due to reflection resonances of the sample and not enhanced
particle-particle interactions mediated by sample waveguide modes.

\acknowledgements

This research was supported by 
the Swedish Natural Science Research Council (NFR) 
and the SSF through the Nanometer Consortium at Lund University.
Discussions with P. Apell and A. Leitner are gratefully acknowledged.

\appendix


\section{Coupling to plane waves}
\label{Appendplane}

\subsection{Plane-wave-to-multipole coupling}

In this Appendix we present results for  the coupling factors 
between multipole fields and plane waves.
For a detalied derivation we refer to Ref.\ \onlinecite{MCD}.

We consider first a plane wave written on the general form,
\be 
  {\bf E}({\bf r})= \left\{ E^{(s)} \hat{s}
 + E^{(p)} \hat{p} \right\}
  e^{i{\bf q}\cdot{\bf r}} 
\label{planewave}
\ee
where 
$\hat{s}=[\hat{z}\crossprod \hat{\bf q}_{\|}]$
and 
$\hat{p} =-[\hat{q} \crossprod (\hat{z}\crossprod\hat{\bf q}_{\|})]$
that  is incident on a sphere.
This wave can be expanded in terms of electric and magnetic multipoles
around the center ${\bf R}_j$ of sphere $j$ as 
\be 
  {\bf E}({\bf r}) = \sum _{lm} k\, a_{lm}^{(M)}\,  j_l(k r)\Xvec_{lm}+
 i\nabla\crossprod
 \left[a_{lm}^{(E)} j_l(kr)\Xvec_{lm}\right],
\label{Eseries}
\ee 
where $r=|{\bf r}-{\bf R}_j|$.

The coefficients in this expression depend linearly on the amplitudes of the 
incoming plane wave, i.e.\
\barr
 &&
 a_{lm}^{(E)} 
= 
 f_{lm}^{Ep}({\bf q}) 
 E^{(p)}
+
 f_{lm}^{Es}({\bf q}) 
 E^{(s)}, \ \ {\rm and}
\breakeq
 a_{lm}^{(M)} 
= 
 f_{lm}^{Mp}({\bf q}) 
 E^{(p)}
+
 f_{lm}^{Ms}({\bf q}) 
 E^{(s)},
\earr
where
\be 
 f_{lm}^{Ep}({\bf q})= f_{lm}^{Ms}({\bf q}) = 
  k^{-1} \, U_{lm} ({\bf q}) \,
   e^{i{\bf q}\cdot{\bf R}_j},
\ee
and
\be 
 f_{lm}^{Es}({\bf q}) = \, -\, 
 f_{lm}^{Mp}({\bf q}) 
=
  k^{-1} \, V_{lm} ({\bf q})\, 
   e^{i{\bf q}\cdot{\bf R}_j}.
\ee
The quantities, $U_{lm}({\bf q})$ and
$V_{lm}({\bf q})$,
are given by
\barr
  U_{lm}({\bf q}) &&= 
  - \frac{2\pi i^l (-1)^m}  {\sqrt{l(l+1)}} 
   \left[
 s_{+}\, F_{+}(l,m)\, Y_{l,-m-1}(\Omega_q)\,
 + \right.
\breakeq
 \left.   +
 s_{-}\, F_{-}(l,m)\, Y_{l,-m+1}(\Omega_q)\,
 \right],
\label{Ulmeq}
\earr
where $\Omega_q$ denotes the direction, 
the angles $\theta_q$ and $\phi_q$, of ${\bf q}$ and 
$$
 s_{\pm} = 
 \hat{s}\cdot\hat{x}
 \pm
 i
 \hat{s}\cdot\hat{y},
$$
and
\barr
  V_{lm}({\bf q}) &&= 
  - \frac{2\pi i^l (-1)^m}  {\sqrt{l(l+1)}} 
   \left[
 \eta_{+}\, F_{+}(l,m)\, Y_{l,-m-1}(\Omega_q)\,
 + \right.
\breakeq
 \left.   +
 \eta_{-}\, F_{-}(l,m)\, Y_{l,-m+1}(\Omega_q)\,
 - \eta_z\, 2m\, Y_{l,-m}(\Omega_q)
 \right],
\label{Vlmeq}
\earr
with
$$
 \hat{\eta}=k^{-1} {\bf q} \crossprod (\hat{z}\crossprod\hat{q}_{\|}); \ \ 
 \eta_{\pm} = \eta_x \pm i \eta_y,
$$
and
\barr
&&
 F_{+}(l,m)= \sqrt{(l-m)(l+m+1)},  \ \ {\rm and}
\breakeq
 F_{-}(l,m)= \sqrt{(l+m)(l-m+1)}.
\label{Fdef}
\earr

The above expressions are also valid for evanescent waves that decay
exponentially in the positive or negative $z$ direction.
The most general wave vector we consider here
lies on the mass shell, and can be written
\be
 {\bf q} = {\bf q}_{\|} \pm \hat{z}\, \sqrt{k^2 - |{\bf q}_{\|}|^2}.
\ee
The plus sign holds for waves that are outgoing 
(propagating or evanescent) in the positive $z$ direction,  and then
$\cos{\theta_q}$ is either real and positive, or lies on the positive
imaginary axis.
The minus sign holds for outgoing waves in the negative $z$ direction.

\subsection{Multipole-to-plane-wave coupling}

We also need to know how much the outgoing waves from a multipole 
contributes to the amplitude of a certain plane wave.
To this end we apply the Kirchoff integrals. 
Thus, for a magnetic 
multipole we calculate ${\bf E}$ from Eq.\ (\ref{KirchE}), 
while for an electric multipole we 
calculate ${\bf B}$ from Eq.\ (\ref{KirchB}).
To carry out the integration we use the following form for the 
Green's function
\be
 G(\rvec,\rpvec)
 =i
 \int \frac{d^2q_\|}{(2\pi)^2}
 \frac{e^{i\sqrt{k^2-q_{\|}^2}|z-z'|}}{2\sqrt{k^2-q_{\|}^2}}
 e^{i{\bf q}_{\|}\cdot(\rvec_{\|}-\rpvec_{\|})}
.
\label{Gcart}
\ee
The scalar plane wave appearing here is expanded in terms of spherical 
harmonics, so that the overlap integrals with the multipole fields
can be carried out.
As is evident from Eq.\ (\ref{Gcart}), the radiated field 
is a linear combination of many plane waves,
and the result can be written as 
\be
  {\bf E}({\bf r})=
 \sum_{lm,\sigma}
 a_{lm}^{\sigma} 
 \int \frac{d^2 q_{\|}}{(2\pi)^2}
 \left[
 g_{lm}^{s\sigma}({\bf q}) \hat{s}
 + g_{lm}^{p\sigma}({\bf q})
  \hat{p}
 \right]
 e^{i{\bf q}\cdot {\bf r}}.
\ee
The coupling factors are found to be
\be
 g_{lm}^{pE}({\bf q}) = 
 g_{lm}^{sM}({\bf q}) = 
 \frac{e^{-i{\bf q}\cdot{\bf R}_j}} { 2 \sqrt{ k^2 - |{\bf q}_{\|}|^2 } }
  (-1)^{l+m+1} U_{l,-m}({\bf q}),
\ee
and
\be
 g_{lm}^{sE}({\bf q}) = 
  -\, \, g_{lm}^{pM}({\bf q}) = 
 \frac{(-1)^{l+m+1}\, e^{-i{\bf q}\cdot{\bf R}_j}} 
 { 2 \sqrt{ k^2 - |{\bf q}_{\|}|^2 } }
 V_{l,-m}({\bf q}).
\ee

\section{Calculation of interparticle coupling}
\label{Appendtdir}

The Fourier transform of $\tensor{t}^{\rm dir}$ is given by
\be
   \tensor{t}_{\bf Q}^{\rm dir} = 
  \sum_{j\neq i} 
  \tensor{t}_{ij}^{\rm dir} e^{-i{\bf Q}\cdot ({\bf R}_i-{\bf R}_j)}
\ee
where $i$ can be any site.
If all spheres in the array sends out waves described by 
$\vec{b} e^{i{\bf Q}\cdot {\bf R}_j}$, the waves received at site $i$
(without any intermediate scattering taking place) is
\barr
   \vec{a}_i 
   =
   \sum_{j\neq i}
   \tensor{t}_{ij}^{\rm dir}
   e^{i{\bf Q}\cdot{\bf R}_j}
   \vec{b}
   &&
   =
   e^{i{\bf Q}\cdot{\bf R}_i}
   \sum_{j\neq i}
   \tensor{t}_{ij}^{\rm dir} \,
   e^{-i{\bf Q}\cdot({\bf R}_i-{\bf R}_j)} \,
   \vec{b}
   =
\breakeq
   =
   e^{i{\bf Q}\cdot{\bf R}_i} \,
   \tensor{t}_{\bf Q}^{\rm dir} \,
   \vec{b}.
\earr
In particular, at the site at the origin,
$$
   \vec{a}_0=\tensor{t}_{\bf Q}^{\rm dir} \, \vec{b}.
$$
We therefore calculate $t_{\bf Q}^{\rm dir}$ by identifying the waves
incident on the sphere at the origin. 

We can simplify the detailed calculation by realizing that outside a source 
sphere the outgoing waves appears to originate from the center of the 
sphere, and likewise, when expanding the waves incident on
the sphere at the origin in different multipoles, this can be done at
any distance from the sphere center. 
In the calculations at hand here we therefore set the radius of the 
source sphere to $r_2$ and that of the receiving sphere to $r_1$ and let
both these radii tend to zero.
The physical size (the radius $R$) of the real spheres is of course
relevant to the physics, but this enters the calculation only through the
response functions $s_l^{(E)}$ and $s_l^{(M)}$.

Consider now an electric multipole source with
angular momentum quantum numbers $l'$ and $m'$.
We calculate the ${\bf B}$ field this generates near the origin
using the Kirchoff integral in Eq.\ (\ref{KirchB}), where
now the last term vanishes, so that
\barr
 {\bf B}({\bf r}_1)=&&
 \sum_{j\neq 0} 
 r_2^2 \int d\Omega_2 
  \left[
        - \frac{ik}{c} (\hat{r}_2\crossprod{\bf E}({\bf r}_2))
	G({\bf r}_1,{\bf r}_2-{\bf R}_j)
\right.
\breakeq 
\left.
        + (\hat{r}_2\crossprod{\bf B}({\bf r}_2))
	\nabla_2 G({\bf r}_1,{\bf r}_2-{\bf R}_j)
\right] 
   e^{i{\bf Q}\cdot{\bf R}_j}.
\label{BfromE}
\earr
Here the phase factor $e^{i{\bf Q}\cdot{\bf R}_j}$ has been included
explicitly, and ${\bf B}({\bf r}_2)$ and ${\bf E}({\bf r}_2)$ should be
evaluated on the sphere at the origin (even if this, due to the condition
$j\neq0$, does not contribute to the sum).  

The summation over lattice sites only involves the Green's function and the
last phase factor, two scalar quantities.  We therefore change the order
between integration and summation and begin by evaluating the lattice sum 
\barr
  S_{\bf Q} 
  &&
  =  
  \sum_{j\neq 0} 
  G({\bf r}_1,{\bf r}_2-{\bf R}_j) \,
  e^{i{\bf Q}\cdot{\bf R}_j}
  =
\breakeq
   =
  \sum_{j\neq 0} 
 \frac{\exp{[ik|{\bf r}_1- {\bf r}_2 - {\bf R}_j|]}}
  { 4 \pi |{\bf r}_1- {\bf r}_2 - {\bf R}_j|} \,
 e^{i{\bf Q}\cdot {\bf R}_j}
\earr
and its gradient $\nabla_2 S_{\bf Q}$.  The calculation uses Ewald
methods borrowed from KKR theory.\cite{Ham} The sum can be expressed both
in terms of a one-center expansion
\be 
 S_{\bf Q} = 
 \sum_{L=0}^{\infty}  \sum_{M=-L}^{L} 
 D_{LM} j_L(kr_{12}) Y_{LM}(\Omega_{12}),
\label{onecenter}
\ee
where $r_{12}$ and $\Omega_{12}$ denote the length and direction of 
the vector ${\bf r}_{12} = {\bf r}_1 - {\bf r}_2$,
and a two-center expansion
\be
S_{\bf Q} = \sum_{lml'm'}
 A_{lml'm'}
 j_l(kr_1) j_{l'}(kr_2) Y_{lm}(\Omega_1) Y_{l'm'}^* (\Omega_2).
 \label{twocenter}
\ee
The two sets of expansion coefficients are 
related through
\be 
  A_{lml'm'} = 4 \pi \, i^{l'-l}
 \sum_{LM} 
 i^L \,  D_{LM}  \, C_{lml'm'}^{LM} ,
\ee
where $C_{lml'm'}^{LM}$ denotes a Gaunt integral
\be
  C_{lml'm'}^{LM} =
 \int d\Omega \,  Y_{lm}^* (\Omega) \, Y_{LM} (\Omega) \, Y_{l'm'} (\Omega).
\ee
The two expansions serve different purposes. When evaluating
${\bf B} $ from Eq.\ (\ref{BfromE}), and then projecting the result onto 
different multipoles at the origin, we will use the two-center
expansion.  
The one-center expansion is useful because using Ewald
methods  $S_{\bf Q} $ can be split into a long-range part
and a short-range part, and the coefficients $D_{LM}$ can be 
calculated rapidly.

There are  three contributions to  $D_{LM}$,
\be
 D_{LM} = D_{LM}^{(1)} + D_{LM}^{(2)} + D_{00}^{(3)} \delta_{L0} .
\ee
The long-range contribution
$D_{LM}^{(1)}$ 
can be expressed as a sum of crystal-rod integrals
\barr
 D_{LM}^{(1)} && = 
  \frac{ 4 \pi i^L}{A_{\rm cell}  \, k^L}
 \sum_{\bf G} 
 \int \frac{dq_z}{2\pi} 
\breakeq \times
 \frac{| {\bf q} + {\bf G} |^L }{| {\bf q} + {\bf G} |^2 - k^2 }  \,
  Y_{LM}^*(\Omega_{{\bf q} + {\bf G}})  \,
 e^{(k^2 - | {\bf q} + {\bf G} |^2)/\eta },
\earr
here ${\bf q} = {\bf Q} + \hat{z} q_z$ and $\eta$ is a separation parameter.
$D_{LM}^{(2)}$ comes from a direct summation over the lattice, and thus
collects short-range interactions
\be
 D_{LM}^{(2)}  = 
 \frac{2^{L+1}}{\sqrt{\pi}\, k^L } 
 \sum_{j\neq 0}
 |{\bf R}_j|^L
  \, e^{i{\bf Q}\cdot{\bf R}_j} \,
  Y_{LM}^* (\Omega_{{\bf R}_j})   \,
  I_L(|{\bf R}_j|, k)
\ee
where the integral
\be 
  I_L(|{\bf R}_j|, k)
  =
 \int_{\sqrt{\eta}/2}^{\infty}
   d \xi \, \xi^{2L} \exp{[ -\xi^2 |{\bf R}_j|^2 + k^2/4\xi^2]}
\ee
can be related to the complementary error function.
The last contribution compensates for the fact that $D_{00}^{(1)}$
contains some self-interaction contributions,
\be 
 D_{00}^{(3)} = 
       - \frac{i k} { \sqrt{4\pi} }
 +
 \frac{\sqrt{\eta}} {2\pi} \sum_{n=0}^{\infty} 
 \frac{ (k^2/\eta)^n  } {n!\, (2n-1)} 
.
\ee
In the calculations here the separation parameter $\eta$ was related to the
lattice parameter $a$ as $\eta=3.24/a^2$; 
the value of $\eta$ of course does not affect the result
for $D_{LM}$.

To proceed with the calculation of $\tensor{t}_{\bf Q}$, 
we insert the two-center expansion Eq.\ (\ref{twocenter}) into 
Eq.\ (\ref{BfromE}), together with expressions for 
$\hat{r}_2\crossprod{\bf E}$ and
$\hat{r}_2\crossprod{\bf B}$ for a certain electric multipole $(L'M')$.
The resulting surface integrations are now straightforward since 
$\Xvec_{L'M'}$ is a linear combination of 
$Y_{L',M'+1}$, $Y_{L',M'}$, and $Y_{L',M'-1}$.
We get
\barr
 &&   {\bf B}({\bf r}_1)  = - \frac{i}{c} \, 
 \frac{ b_{L'M'}^{(E)} } {2 \sqrt{L'(L'+1)} }\,\,
 \sum_{lm} 
 j_l(kr_1) Y_{lm}(\Omega_1)
\breakeq
\times
  \left\{
 \hat{n}_{-} F_{+}(L',M') \, A_{l,m,L',M'+1}
\right.
\breakeq
+
 \hat{n}_{+} F_{-}(L',M')  \, A_{l,m,L',M'-1}
+
 \hat{n}_{z} 2M'\, A_{l,m,L',M'}
 \left.  \right\}.
\earr
In this expression 
$ \hat{n}_{\pm} = \hat{x} \pm i \hat{y}$, $\hat{n}_z = \hat{z}$,
and $F_+$ and $F_-$ are defined in Eq.\ (\ref{Fdef}).

Next we project ${\bf B}$ onto different multipoles around the origin.
The field corresponding to an electric multipole $(LM)$ can be written
$$
  {\bf B} = \frac{k}{c} a_{LM}^{(E)} j_L(kr_1) \Xvec_{LM}(\Omega_1),
$$
therefore
\be
   \frac{k}{c} a_{LM}^{(E)} j_L(kr_1) = 
    \int d \Omega_1 (\Xvec_{LM}(\Omega_1))^* \cdot {\bf B}({\bf r}_1),
\ee
and the final result for 
$t_{LML'M'{\bf Q}}^{(EE)} = a_{LM}^{(E)}/b_{L'M'}^{(E)} $
reads
\barr 
  && t_{LML'M'{\bf Q}}^{(EE)} 
   = - \frac{i}{2k}\, 
 \frac{1}{\sqrt{L(L+1)L'(L'+1)}}
\breakeq
\times
  \left\{
   F_{+}(L,M)\, F_{+}(L',M')
   \, A_{L,M+1,L',M'+1}
 \right.
\breakeq
+
   F_{-}(L,M)\, F_{-}(L',M')
   \, A_{L,M-1,L',M'-1}
\breakeq
 \left.
+
 2MM'\, A_{L,M,L',M'}
 \right\}.
\earr

The corresponding calculation for the projection onto a magnetic multipole 
$(LM)$ is somewhat more involved.  We use the fact that for a magnetic
multipole
$$
 {\bf r}_1\cdot{\bf B}({\bf r}_1) = 
 \frac{1}{c} a_{LM}^{(M)} j_L(kr_1) \sqrt{L(L+1)} Y_{LM}(\Omega_1),
$$
and consequently
\be
 \frac{1}{c} a_{LM}^{(M)} j_L(kr_1) \sqrt{L(L+1)} 
 =
 \int d \Omega_1 \, 
 Y_{LM}^*(\Omega_1)  \left[ {\bf r}_1 \cdot {\bf B} \right].
\ee
Here the Cartesian components of ${\bf r}_1$ introduce spherical harmonics
$Y_{11}$, $Y_{10}$,  and $Y_{1-1}$ into the integral. The final result thus
contains Gaunt integrals,
\barr 
  && t_{LML'M'{\bf Q}}^{(ME)} 
   = - \frac{i}{2k}\, 
 \frac{2L+1}{\sqrt{L(L+1)L'(L'+1)}}\,
 \sqrt{\frac{4\pi}{3}}\,
\breakeq
\times
  \left\{
   \sqrt{2}\, C_{LM,L-1,M+1}^{1,-1}\, F_{+}(L',M')
   \, A_{L-1,M+1,L',M'+1}
 \right.
\breakeq
-
   \sqrt{2}\, C_{LM,L-1,M-1}^{1,1}\, F_{-}(L',M')
   \, A_{L-1,M-1,L',M'-1}
\breakeq
 \left.
+
 C_{LM,L-1,M}^{1,0}\, 2M'\, A_{L-1,M,L',M'}
 \right\}.
\earr

We have determined half of the elements in $\tensor{t}_{\bf Q}$.  If the
sources instead are magnetic multipoles we use the other Kirchoff integral
in Eq. (\ref{KirchE}) to calculate ${\bf E}({\bf r}_1)$ around the origin.
The calculations are perfectly analogous to the ones above and in the end
one finds the following symmetry of $\tensor{t}$,
\barr
   && 
    t_{LML'M'{\bf Q}}^{(MM)}  = t_{LML'M'{\bf Q}}^{(EE)} , \  \ \ {\rm and} 
\breakeq
   t_{LML'M'{\bf Q}}^{(EM)}  = -  \, t_{LML'M'{\bf Q}}^{(ME)}.
\earr

\section{Calculation of sample-mediated interactions}
\label{Appendtsub}

Here we will calculate the sample-mediated self-interaction $\tensor{w}$
as well as $\tensor{t}_{\bf Q}^{\rm sub}$.

Starting with $\tensor{w}$,
suppose a certain multipole $(l'm'\sigma')$ is excited on a sphere.
Then plane waves are sent out from the sphere, reflected off the sample
surface either as s or p polarized light, and after that the wave
impinges on the sphere again.  The strength $a_{lm}^{\sigma}$ of the incident
wave can be calculated using the previously introduced functions $f$
and $g$, and the result reads
\barr
  && w_{lml'm'}^{\sigma \sigma'} =
 \sum_{\sigma''} \int \frac{d^2q_{\|}}{(2\pi)^2}
\breakeq
\times
 f_{lm}^{\sigma\sigma''}  ({\bf q_{+}}) \chi_{\sigma''}({\bf q}_{\|}) 
 g_{l'm'}^{\sigma''\sigma'} ({\bf q}_{-})
 e^{2i\sqrt{k^2-q_{\|}^2} |z_0|},
\earr
where 
$$
{\bf q}_{\pm} = {\bf q}_{\|} \pm \hat{z} \sqrt{k^2 - |{\bf q}_{\|}|^2}
$$
and $\sigma''$ denotes the plane-wave polarization ($s$ or $p$).
The angular integration (over the different directions of ${\bf q}_{\|}$)
is straightforward, and yields a finite result only when $m=m'$.
The remaining integration over $|{\bf q}_{\|}|$ must be done numerically.

Once $\tensor{w}$ is known, it is relatively easy to calculate 
$\tensor{t}_{\bf Q}^{\rm sub}$.  
If the same multipole is excited on all the spheres,
with relative phase $e^{i{\bf Q}\cdot {\bf R}_j}$, the incident wave at the 
origin must be proportional to a sum  of $\tensor{t}_{\bf Q}^{\rm sub}$ and 
$\tensor{w}$.
This sum  is easily calculated thanks to the periodicity of
the radiating system.  A certain tensor element is found through a summation
over the 2D reciprocal lattice
\barr
 && t_{lml'm',{\bf Q}}^{{\rm sub},{\sigma\sigma'}} 
 + 
 w_{lml'm'}^{\sigma\sigma'} =  
  \frac{1}{A_{\rm cell}} \sum_{\sigma''} \sum_{\bf G}  
\breakeq
\times
 f_{lm}^{\sigma\sigma''}  ({\bf q}_{+}) \chi_{\sigma''}({\bf q}_{\|}) 
 g_{l'm'}^{\sigma''\sigma'} ({\bf q}_{-})
 e^{2i\sqrt{k^2-q_{\|}^2} |z_0|},
\earr
where now 
$$
{\bf q}_{\|}={\bf Q} +{\bf G},
\ \ {\rm and} \ \ 
 {\bf q}_{\pm} = {\bf q}_{\|} \pm \hat{z} \sqrt{k^2 - |{\bf q}_{\|}|^2}.
$$

\end{multicols}

\medskip
\begin{figure}
\centerline{
\psfig{file=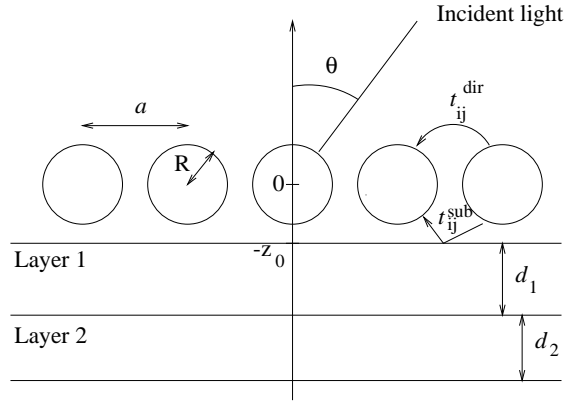,width=7.5cm,angle=0}
}
\vspace{0.1 cm}
\caption{
Schematic illustration of the sample and overlayer of spheres.
The two sample layers have thicknesses $d_1$ and $d_2$, respectively.  
A lattice (lattice gas) of spherical particles with radius $R$ 
and lattice parameter $a$ covers the sample. The particles interact through
direct $\tensor{t}_{ij}^{\mathrm dir}$ and sample-mediated 
$\tensor{t}_{ij}^{\mathrm sub}$ interactions.
(The empty space between the particle layer and the sample is there only for
illustrative purposes; the particles do rest on the sample.)
}
\label{Fig1}
\end{figure}

\medskip
\begin{figure}
\centerline{
\psfig{file=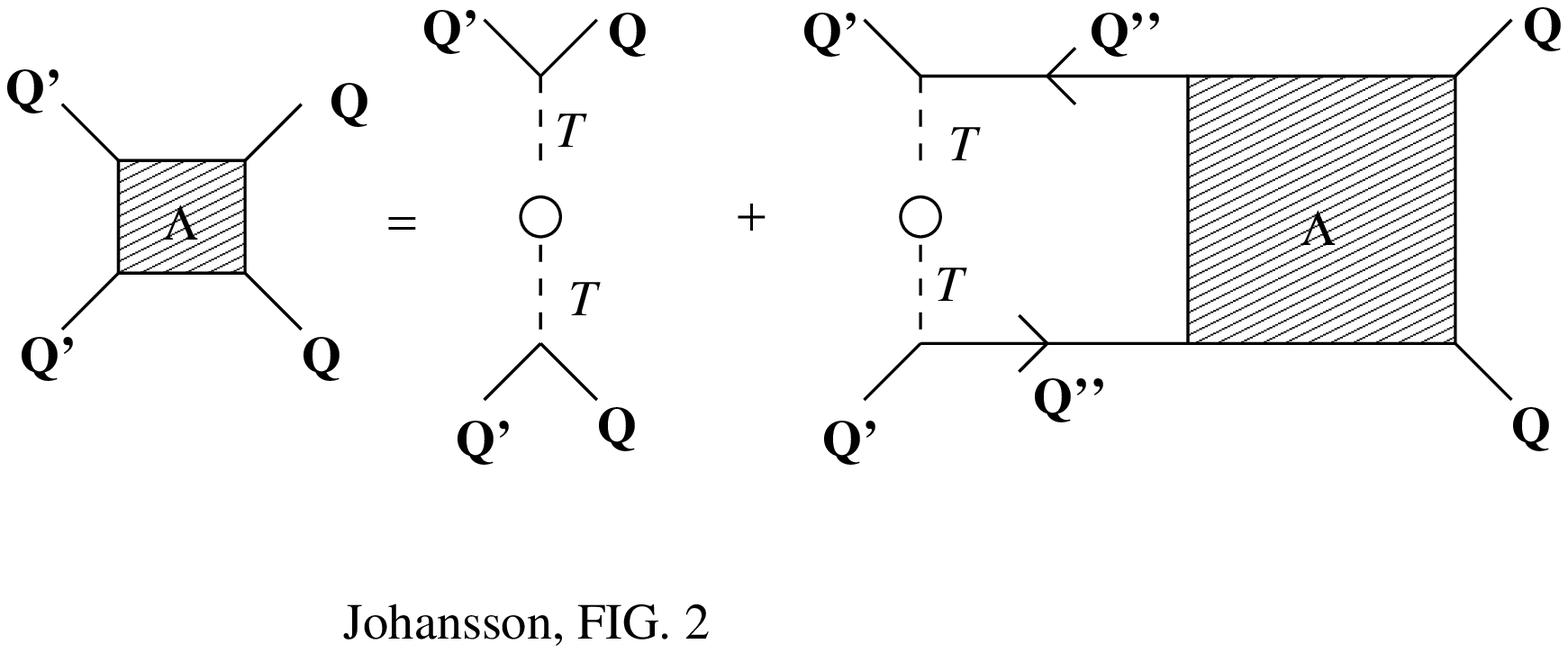,width=7.5cm,angle=0}
}
\vspace{0.1 cm}
\caption{
   Diagrammatic representation of the Bethe-Salpeter equation illustrating
   how the vertex function corresponding to the diffuse scattering intensity 
   is built up by  contributions from a single
   scattering event (first term) as well as by contributions from repeated
   scattering events.
}
\label{Bethefig}
\end{figure}

\newpage
\medskip
\begin{figure}
\centerline{
\psfig{file=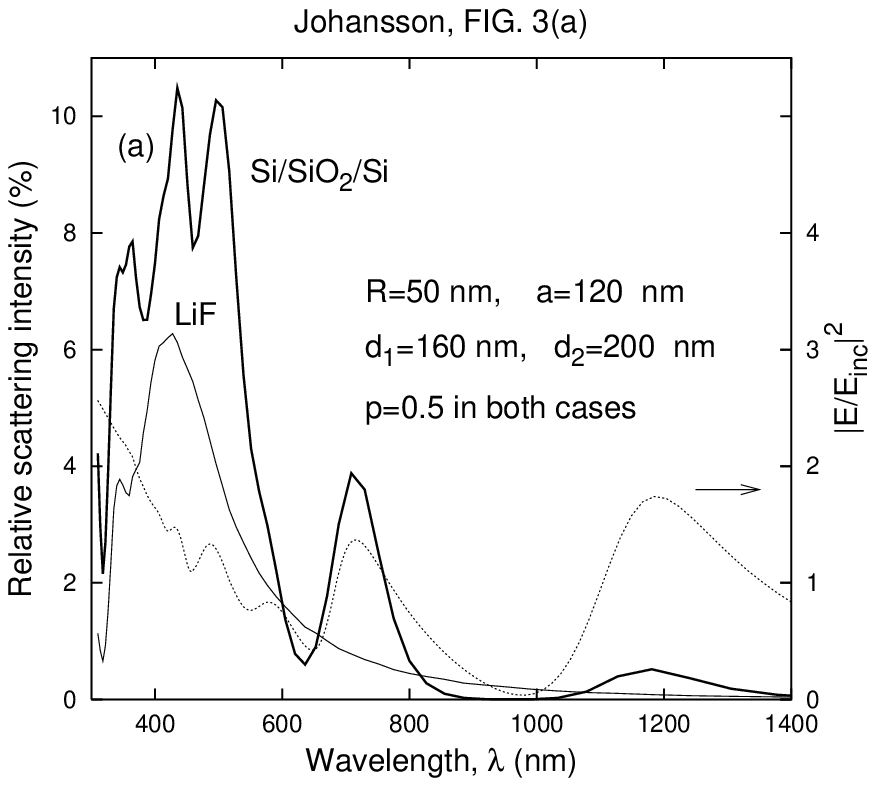}
}
\vspace{0.1 cm}
\centerline{
\psfig{file=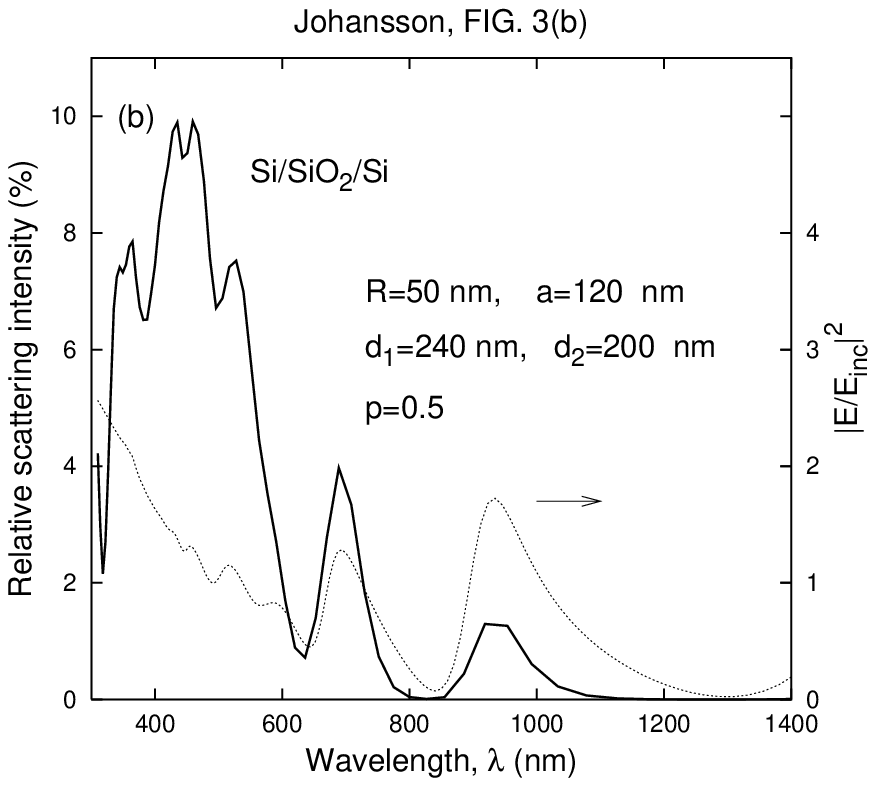}
}
\vspace{0.1 cm}
\caption{
(a)
Relative intensity of the diffusely {\em reflected} light 
(integrated over the full solid angle $2\pi$) from a disordered 
overlayer of Ag spheres. The sample is illuminated at normal incidence.
The two solid curves show 
results obtained with a layered Si/SiO$_{2}$/Si sample and
with a homogeneous LiF sample, respectively. 
The dotted curve shows the magnitude (squared) of the driving electric field 
relative to
the incident field  in the plane of the sphere centers.
This field  was calculated in the presence of the 
Si/SiO$_{2}$/Si sample, but without the Ag particles.
(b)
The corresponding results obtained with 
an increased thickness of the top Si layer.
The maxima in the driving field have shifted to larger $\lambda$
[the 1200-nm-peak from  (a) falls outside the figure at $\approx $ 1600 nm]
and the scattering-intensity maxima follow.
}
\label{lambdafig}
\end{figure}

\medskip
\begin{figure}
\centerline{
\psfig{file=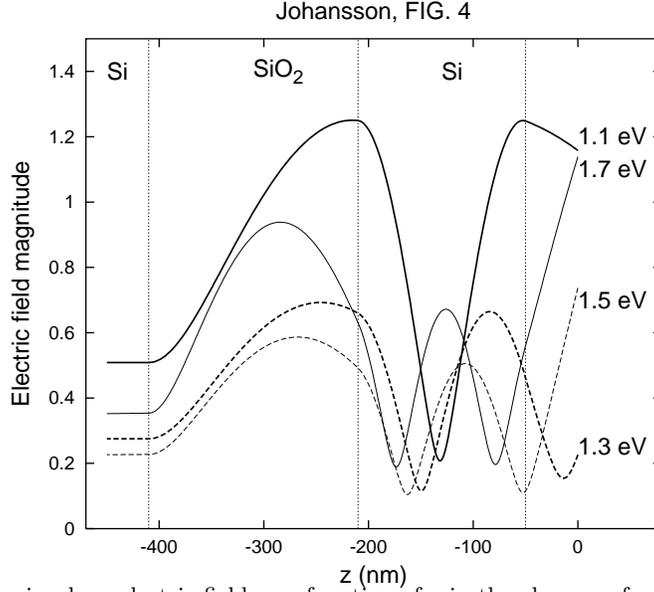}
}
\caption{
The magnitude of the in-plane electric field as a function of $z$
in the absence of any particle overlayer
when a wave with amplitude 1 impinges at normal incidence 
on a Si/SiO$_2$/Si sample with the
same geometric parameters as in Fig.\ \protect\ref{lambdafig} (a).
Results are shown for four different photon energies as indicated next to
the curves.  The corresponding free-space wavelengths are
1130 nm,  950 nm, 825 nm, and 730 nm, respectively.
Note that the value of the field magnitude at $z=0$, the sphere center,
equals the driving field whose square is plotted in 
Fig.\ \protect\ref{lambdafig} (a).
}
\label{fieldfig}
\end{figure}

\medskip
\begin{figure}
\centerline{
\psfig{file=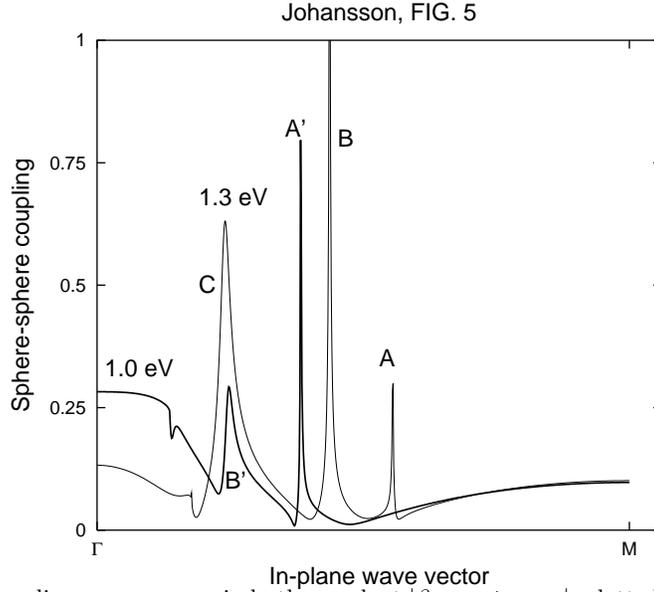}
}
\caption{
The sphere-sphere coupling, or more precisely the product 
$|\beta_{11,11{\bf Q}} t_{11,11{\bf Q}}|$,
plotted for two different
photon energies 1 eV (thick curve) and 1.3 eV (thin curve) 
as a function of the in-plane wave vector ${\bf Q}$
along the diagonal of the Brillouin zone
from the center point ($\Gamma$) where ${\bf Q}=0$ to the corner point ($M$)
where ${\bf Q}=(2\pi/a)(\hat{x}+\hat{y})$.
The peaks marked A and A' are due to s polarized waveguide modes the ones
marked B and B' are due to p polarized waveguide modes.
The curves exhibit discontinuities (coming from the interaction $t$) 
at the point where $|{\bf Q}|=k$.
}
\label{couplfig}
\end{figure}

\medskip
\begin{figure}
\centerline{
\psfig{file=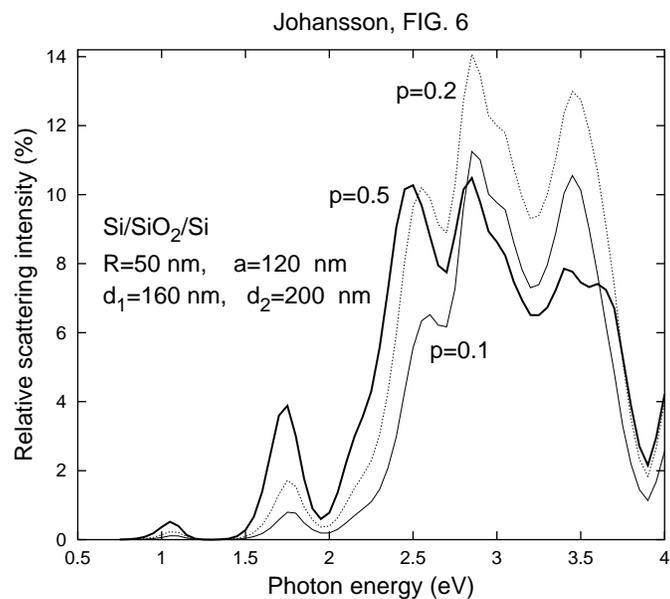}
}
\vspace{0.1 cm}
\caption{
Diffusely reflected light intensity from a Si/SiO$_2$/Si sample 
when the coverage $p$ of Ag spheres is varied as
indicated next to the curves. The other parameter values are the same as in
Fig.\ \protect\ref{lambdafig} (a).
}
\label{coveragefig}
\end{figure}

\medskip
\begin{figure}
\centerline{
\psfig{file=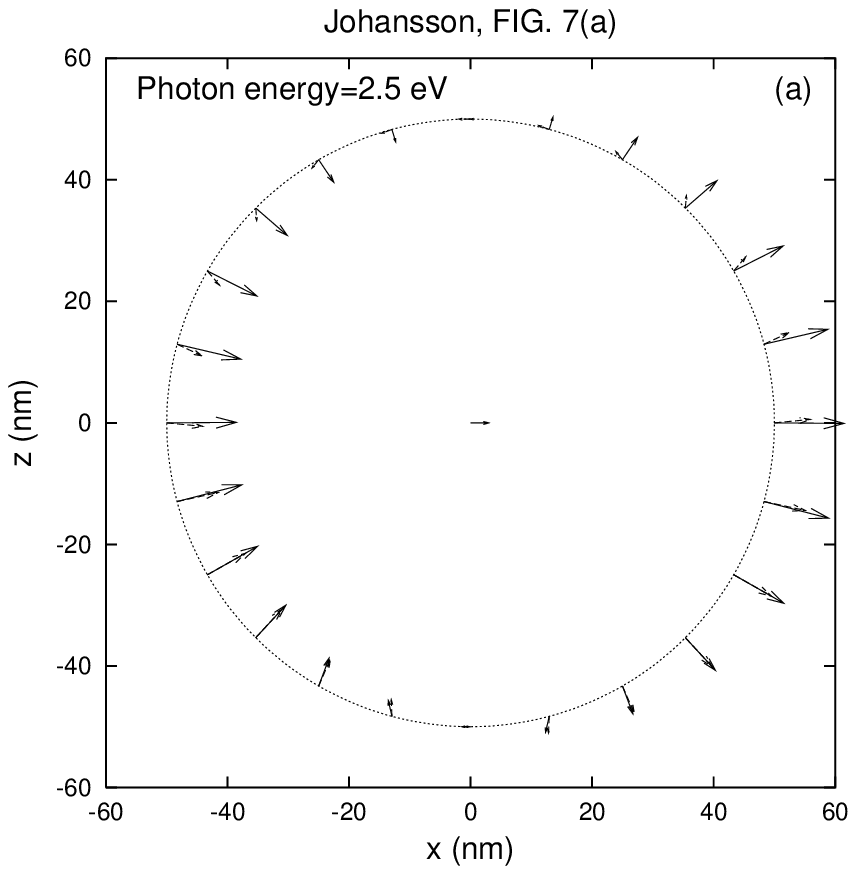}
}
\vspace{0.1 cm}
\centerline{
\psfig{file=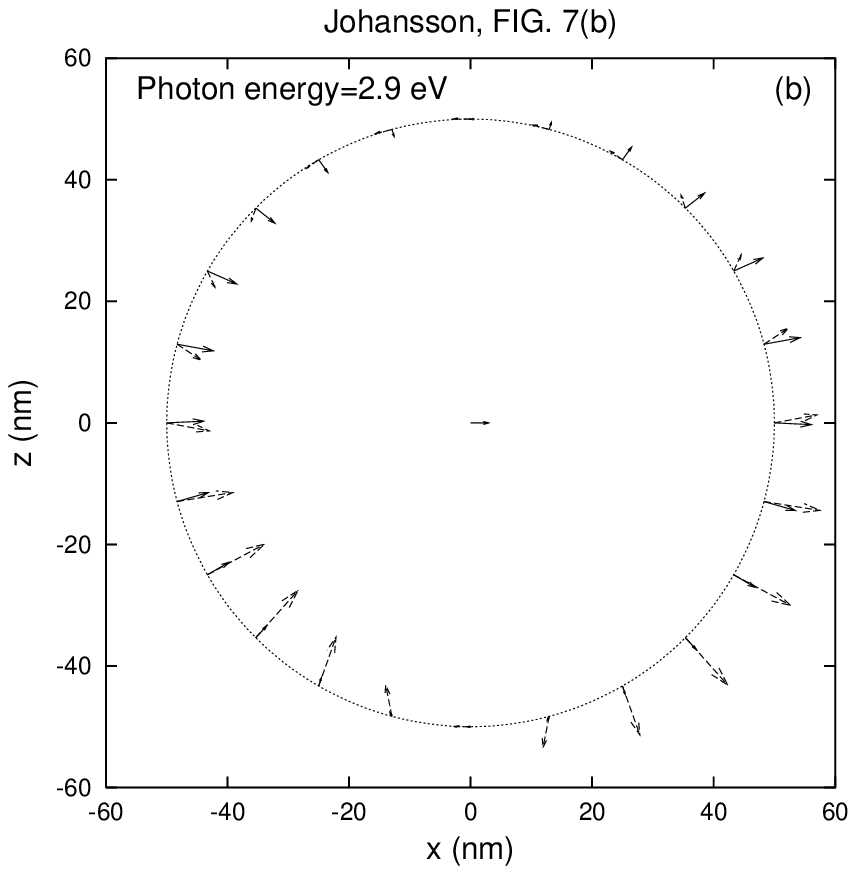}
}
\vspace{0.1 cm}
\centerline{
\psfig{file=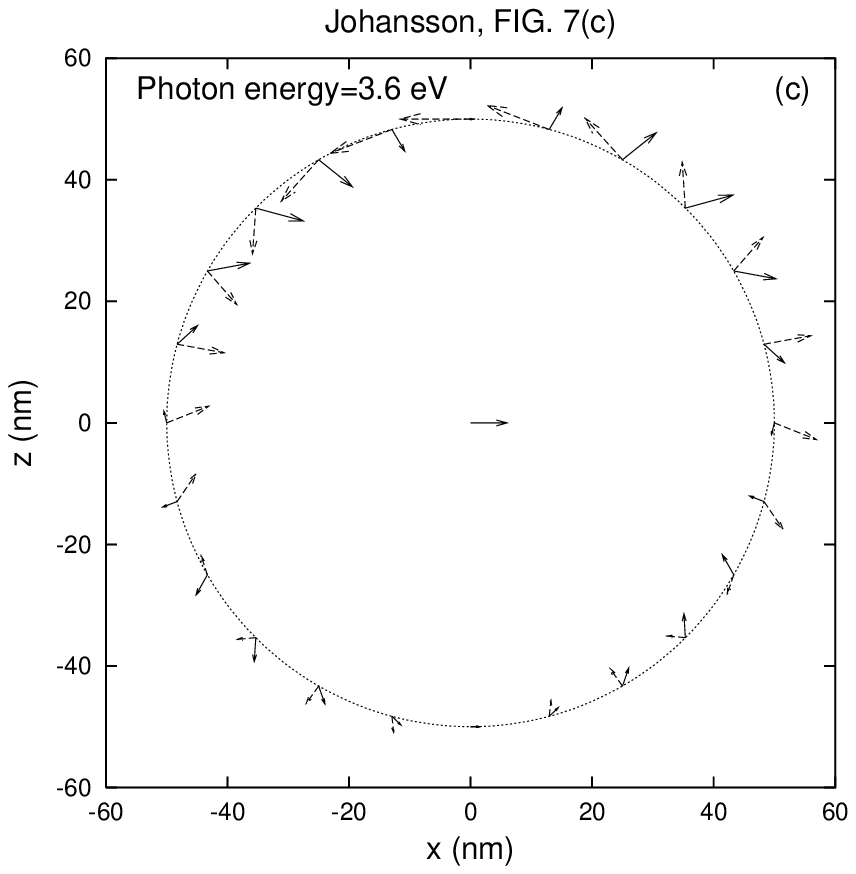}
}
\vspace{0.1 cm}
\caption{
The fields around an average scatterer for three different photon energies:
(a) 2.5 eV, (b) 2.9 eV, and (c) 3.6 eV.
The parameter values are the same as in Fig.\ \protect\ref{lambdafig} (a).
The full arrows give the real part of the field (in phase with the incident
wave) and the broken arrows the imaginary part.
The arrow in the middle of each figure shows the strength of the
field of the incident wave [thus, the scaling used in panel (c) differs
from that of the other panels].
}
\label{vecfig}
\end{figure}

\medskip
\begin{figure}
\centerline{
\psfig{file=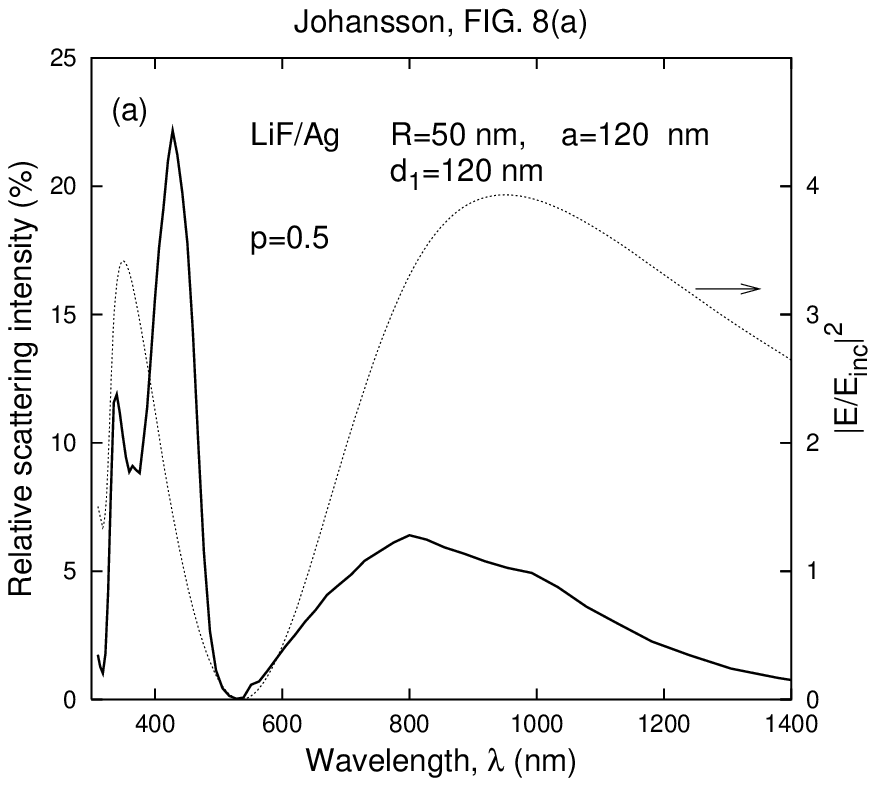}
}
\vspace{0.1 cm}
\centerline{
\psfig{file=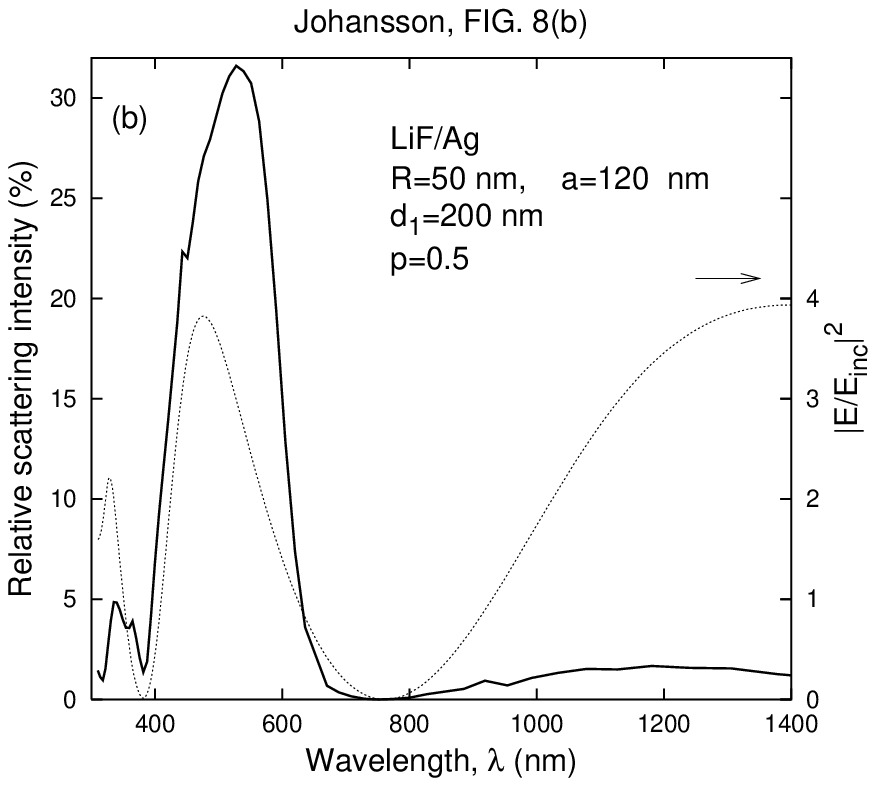}
}
\vspace{0.1 cm}
\caption{
Relative intensity of the diffusely {\em reflected} light 
(integrated over the full solid angle $2\pi$) from a disordered 
overlayer of Ag spheres with a relative coverage $p=0.5$ 
on a LiF/Ag sample. 
In both plots the full curves show the scattering intensity, whereas the
dotted curves show the result for the (squared) driving field.
The thickness of the LiF layer
is (a) 120 nm and (b) 200 nm, respectively, in the two plots.
}
\label{silverfig}
\end{figure}


\begin{references}
\bibitem{lmaxnote}
In the numerical calculations we used $l_{max} =3$. This is sufficient 
to obtain convergence for the scattered intensity. 
However, the results for the fields near a particle presented in 
Fig.\ \protect\ref{vecfig} would change to a certain extent if more multipoles
were included.
\bibitem{expres} 
   In the experiment the peaks occur at about 780 nm and 1250 nm. The
   sample used there had also a 30 nm thick LiF top layer.
\end{references}
\end{document}